\documentclass[fleqn,usenatbib]{mnras}

\usepackage[T1]{fontenc}

\DeclareRobustCommand{\VAN}[3]{#2}
\let\VANthebibliography\thebibliography
\def\thebibliography{\DeclareRobustCommand{\VAN}[3]{##3}\VANthebibliography}

\usepackage{graphicx}	
\usepackage{amsmath}	
\usepackage{amssymb}	
\usepackage{multirow}
\usepackage{cuted}
\usepackage{hyperref}
\usepackage{url}
\usepackage[dvipsnames]{xcolor}
\usepackage{soul}
\usepackage{todonotes}

\usepackage[export]{adjustbox}

\newcommand{\kms}{\mbox{$\mathrm{km\,s}^{-1}$}}
\newcommand{\Msy}{\mbox{$\mathrm{M_{\sun}\,yr}^{-1}$}}
\newcommand{\Teff}{\mbox{$T_\mathrm{eff}$}}
\newcommand{\Msun}{\mbox{$\mathrm{M}_{\sun}$}}
\newcommand{\Rsun}{\mbox{$\mathrm{R}_{\sun}$}}

\newcommand{\Line}[3]{#1\,{\sc #2}~$\lambda$#3}

\usepackage{newtxtext,newtxmath}

\title[ASAS\,J071404+7004.3~--~a close, bright nova-like with gusty winds]{ASAS\,J071404+7004.3~--~a close, bright nova-like cataclysmic variable with gusty winds}

\author[K. Inight et al.]{
K.~Inight,$^{1}$\thanks{E-mail: keith.inight@gmail.com }
B.~T.~G\"ansicke,$^{1}$
D.~Blondel,$^{1}$
D.~Boyd$^{2},$
R.~P.~Ashley,$^{3}$
C.~Knigge,$^{6}$
\newauthor
K.~S.~Long,$^{7}$
T.~R.~Marsh,$^{1}$
J.~McCleery,$^{1}$
S.~Scaringi,$^{4}$
D.~Steeghs,$^{1}$
J.~R.~Thorstensen,$^{8}$
\newauthor
T.~Vanmunster,$^{5}$
P.~J.~Wheatley$^{1}$
\\
$^{1}$Department of Physics, University of Warwick, Coventry, CV4 7AL, UK\\
$^{2}$BAA Variable Star Section, West Challow Observatory, OX12 9TX, UKK\\
$^{3}$Isaac Newton Group of Telescopes, Apartado de Correos 321, E-38700 Santa Cruz de La Palma, Spain\\
$^{4}$Centre for Extragalactic Astronomy, Department of Physics, Durham University, DH1 3LE, UK\\
$^{5}$CBA Belgium Observatory \& CBA Extremadura Observatory, Walhostraat 1a, B-3401 Landen, Belgium\\
$^{6}$School of Physics and Astronomy, University of Southampton, Highfield, Southampton, SO17 1BJ, UK\\
$^{7}$Space Telescope Science Institute, 3700 San Martin Drive, Baltimore, MD, 21218, USA\\
$^{8}$Department of Physics and Astronomy, Dartmouth College, Hanover NH 03755, USA
}

\date{Accepted XXX. Received YYY; in original form ZZZ}

\pubyear{2020}

\begin{document}
\label{firstpage}
\pagerange{\pageref{firstpage}--\pageref{lastpage}}
\maketitle

\begin{abstract}
Despite being bright ($V\simeq11.8$) and nearby ($d=212$\,pc) ASAS\,J071404+7004.3 has only recently been identified as a nova-like cataclysmic variable. We present time-resolved optical spectroscopy obtained at the Isaac Newton and the Hiltner and McGraw-Hill Telescopes, together with \textit{Swift} X-ray and ultraviolet observations. We combined these with \textit{TESS} photometry and find a period of 3.28\,h and a mass transfer rate of $4-9\times 10^{-9}\,\Msy$.  Historical photometry shows at least one low state establishing the system as a VY\,Scl star. Our high-cadence spectroscopy also revealed rapidly changing winds emanating from the accretion disc. We have modelled these using the Monte Carlo  \textsc{python} code and shown that all the emission lines could emanate from the wind~--~which would explain the lack of double-peaked lines in such systems. In passing, we discuss the effect of variability on the position of cataclysmic variables in the \textit{Gaia} Hertzsprung-Russell diagram. 

\end{abstract}



\section{Introduction}
Cataclysmic variables (CVs) are short-period
($\simeq1-10\mathrm{h}$) binary systems consisting of a white dwarf accreting from a main sequence companion \citep{1995cvs..book.....W}. In CVs with non-magnetic (or very weakly magnetic) white dwarfs, the mass transfer proceeds via an accretion disc. Below a critical mass transfer rate ($\dot{M}\simeq1-5\times 10^{-9}$\,\Msy) thermal instabilities in accretion discs result in dwarf novae outbursts, in which the disc cycles between a cool and a hot state. CVs above the critical $\dot{M}$ appear as nova-likes which have steady, hot and luminous discs \citep{1984A&A...132..143M,1986ApJ...305..261S,1988ApJ...333..227C}. In reality, CVs display a range of behaviour that goes beyond the simple dichotomy described above. Some CVs have mass transfer rates close to the critical value, and cycle between undergoing regular outbursts and states where their discs are in a steady hot state  \citep{2001A&A...369..925B}. Even though the accretion discs in nova-likes are in a (relatively) steady luminous state, they can exhibit small ``stunted'' outbursts \citep{ 1998AJ....115.2527H}.  In addition, some nova-likes occasionally enter a low state during which mass loss from the donor drops significantly, or even shuts off completely \citep{ 2004AJ....128.1279H}. During this time they may exhibit dwarf nova outbursts \citep{1998ApJ...499..348K}. 

In summary, CVs provide an ideal environment for the  study of accretion processes under a range of physical conditions, with applications to a range of other systems including young stellar objects \citep{1987IAUS..115....1L} and quasars \citep{1999qagn.book.....K}. Key areas where our understanding of CV accretion discs is incomplete are the nature of the viscosity driving the mass transfer through the disc \citep{1973A&A....24..337S}, the detailed physical structure of the disc \citep{2017ApJ...846...52G} and of the interface between the disc and the white dwarf (the so-called ``boundary layer'', see Section 2.5.4 in \citealt{2003cvs..book.....W}). In addition, CVs often display signatures of winds \citep{1997cwh..conf..111M}, and in some cases jets \citep{ 2020NewAR..8901540C}, and both the structure and the driving mechanisms of these are not very well established. 

Discovering new, bright examples of CVs remains important to grow the arsenal of diagnostic tools that are available to investigate these topics. However, because of their steady discs, nova-like variables have remained difficult to identify, in contrast with dwarf novae that stand out in photometric time-domain surveys because of their large-amplitude variability \citep{2005ASPC..330....3G,  2014MNRAS.441.1186D, 2014MNRAS.443.3174B, 
2020AJ....159..198S}. 

Here we report a detailed study of the bright and nearby nova-like ASAS\,J071404+7004.3 (hereafter ASAS\,J0714+7004). ASAS\,J0714+7004 was identified as the optical counterpart to the X-ray source 1RXS\,J071404.0+700413, and a potential variable star, by \citet{2013AcA....63...53K} who suggested it might be a CV with a putative period of 22.4\,d. ASAS\,J0714+7004  came to our attention as part of a project to identify white dwarf binary candidates located between the main sequence and the white dwarf cooling sequence, using a combination of \textit{Gaia} and \textit{GALEX} photometry and astrometry (Inight et al. in prep).

We summarise our observations and provide a qualitative description of the system in Section\,\ref{sec:obs}, analyse these observations to extract the physical parameters of the system in Section \ref{sec:analysis} and then discuss the implications (Section \ref{sec:discussion}) and conclusions (Section \ref{sec:conclusions}).

\begin{figure*} 
\includegraphics[width=\textwidth]{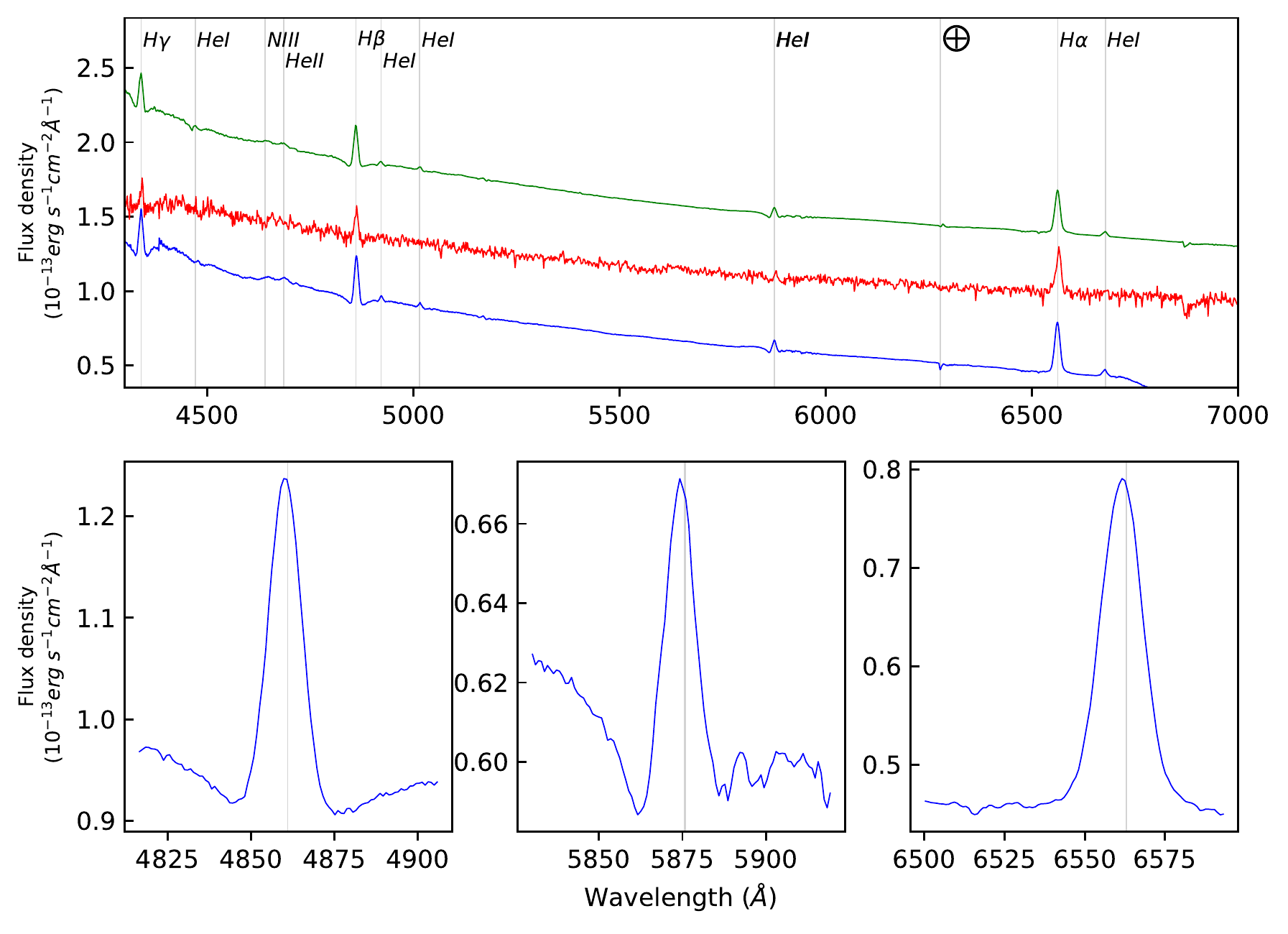}
\caption{\label{fig:average_spectrum}  
Spectra of ASAS\,J071404+7004.3. Top panel: The first spectrum (red~--~offset by 0.6 flux units) was obtained on 2020 February 6.
The second spectrum (blue) is  the average of the INT observations. The third spectrum (green~--~offset by 1.0 flux units) is the average of the MDM observations.
Bottom panel: The average INT spectra for H$\beta$ (left), \ion{He}{i} 5876\,\AA\  (centre) and H$\alpha$ (right). The H$\beta$ line shows an emission line sitting in a broad absorption line, the \ion{He}{i} 5876\,\AA\ triplet line shows a P\,Cygni profile  and the H$\alpha$ line is purely in emission although showing slight traces of a P\,Cygni profile.}
\end{figure*}

\section{Observations}\label{sec:obs}

\subsection{Spectroscopy}

We obtained an identification spectrum on 2020 February 6 using a 280\,mm Schmidt Cassegrain telescope (SCT) equipped with a Shelyak LISA spectroscope, giving a spectral resolution of $R\simeq 1000$. The spectrum was integrated for 30\,min before being stopped by clouds which limited the signal-to-noise ratio. The data were corrected with bias, dark and flat frames, wavelength calibrated using the integrated ArNe lamp, and corrected for the instrumental response function and atmospheric losses using the spectrum of a spectrophotometric standard star taken at the same altitude immediately prior to the target. Processing was performed using the \textsc{isis} spectral analysis software\footnote{\url{http://www.astrosurf.com/buil/isis-software.html}}. The spectrum was then calibrated in absolute flux using a concurrently recorded $V$ magnitude observation using the method described on the BAA website\footnote{
\url{https://britastro.org/sites/default/files/absfluxcalibration.pdf}}.
The SCT spectrum (Fig.\,\ref{fig:average_spectrum}) exhibits a blue slope in the continuum with broad and shallow H$\beta$ and H$\gamma$ absorption lines filled in by single-peaked Balmer emission lines, and additional emission lines from \ion{He}{i}, unambiguously identifying this system as a CV. With a median magnitude of $V\simeq11.8$, ASAS\,J0714+7004 is among the $\simeq20$ brightest CVs known \citep{2003A&A...404..301R}.

ASAS\,J0714+7004 was then observed using the Intermediate Dispersion Spectrograph (IDS) mounted on the 2.54\,m Isaac Newton Telescope (INT) on the island of La Palma, using the R632V grating at a central wavelength of 5720\,\AA. This setup covers the wavelength range 4383--6704\,\AA\ at a dispersion of 0.89\,\AA/pixel. A total of 264 120\,s exposures were taken over a period of 13 days in February 2020 (hereafter the February INT data) and a further long sequence on 2020 December 13 (hereafter the December INT data)~--~see Table~\ref{tab:journal_observations} for details.  

All of the INT spectra were optimally extracted and reduced using the {\sc pamela} and {\sc molly} reduction software \citep{1989PASP..101.1032M}. In order to perform the flux calibration we fitted a spline to the continua of spectrophotometric standard stars taken on the same night as the observations. We then used their published flux as a reference, see e.g. \citet{1990ApJ...357..621M}. For the wavelength calibration, Cu+Ne and Cu+Ar arcs were observed at the start and end of each observation, and approximately once every hour in between during the longer sequences. As a final step in wavelength calibration we extracted the sky spectrum and measured the deviation of the telluric \ion{O}{i} emission line at 5577.4\,\AA\ from its laboratory wavelength and shifted our science spectra to align with this feature. 

Additional spectra of ASAS\,J0714+7004 were obtained prior to the SCT and ING observations\footnote{Upon posting the submitted version of this paper to arXiv, John Thorstensen reminded the other authors that he had observed this star already in 2013.} with the 1.3\,m McGraw-Hill and 2.4\,m Hiltner telescopes at the MDM Observatory on Kitt Peak in Arizona; Table~\ref{tab:journal_mdmobservations} gives a journal. These data will be referred to as the MDM spectra hereafter. On both telescopes we used the modular spectrograph (modspec), which covered from 4210 to 7560\,\AA\ with a resolution of about 4\,\AA\ (full-width at half maximum). To calibrate the wavelengths, we took spectra of Hg, Ne, and Xe lamps in twilight, and derived zero-point shifts to the solution using the night-sky background lines.  We flux-calibrated the spectra using observations of standard stars; experience suggests that the zero point of these calibrations is accurate to $\simeq30$ per cent.  In addition, the continua in modspec data sometimes show unphysical distortions, which tend to average out.

The average of the 264 INT spectra (Fig.\,\ref{fig:average_spectrum}) is very similar to the SCT identification spectrum and the MDM average spectrum, although it has a higher spectral resolution and signal-to-noise ratio. It contains strong single-peaked Balmer emission lines embedded within broad and shallow absorption lines. Overall, the morphology of the spectrum is that of a nova-like, with the absorption arising from an optically thick disc  \citep{1996ASSL..208....3D,1985A&A...151..157H,1993PASP..105...51R}. Unlike dwarf novae, single-peaked Balmer lines are not indicative of low inclination \citep{1991AJ....102..272T, 2007MNRAS.374.1359R, 2011MNRAS.410..963N}. Strong emission lines in nova-likes are in fact  associated with high inclination (see e.g. Fig.\,1 in \citealt{2015MNRAS.450.3331M}). This is because they are believed to originate in the disc winds and hence appear strongest relative to  the underlying disc emission at higher inclinations.  The \ion{He}{i} 5876\,\AA\ line exhibits a P\,Cygni profile.

\begin{table*}
\caption{\label{tab:ephemera} Stellar parameters of ASAS\,J0714+7004.}
\centering
\begin{tabular}{ l lr@{$\,\pm\,$}l r| } 
\hline
\multicolumn {2}{l}{Parameter} & 
\multicolumn {2}{
c}{Value} & References \\
\hline
RA     & $\alpha$\,[h:m:s]         &07:14:04.654486& 0.00001& 1\\

Dec  & $\delta$\,[d:m:s]         & +70:04:18.400755& 0.00001    & 1\\

\textit{Gaia} ERD3 source\_id &          & \multicolumn {2}{c}{1109608206832496512} &  1\\
Parallax               & $\varpi$\,[mas]         & 4.703  &0.02    & 1\\
Distance               & $d$\,[pc]               & 210.15  & 1.0   & 2\\
Apparent magnitude& $G$\,[mag]            & 11.814&0.01	&  1\\
Absolute magnitude & $M_G$\,[mag]            & 5.20&0.016	&  1,2,4\\
Proper motion          & $\mu_\alpha$\,$\mathrm{[mas\, yr^{-1}]}$  & 26.007 & 0.02& 1\\	
                       & $\mu_\delta$\,$\mathrm{[mas\, yr^{-1}]}$  & $-$36.923& 0.025   & 1\\

Mass transfer rate& $\dot{M}$\, [$ 10^{-9}\Msy$]        & \multicolumn {2}{c}{4-9}& 3\\

Period   & $P_{\textrm{orb}}$\,[h]  & 3.2794894&0.0000008 &3\\
Epoch & HJD [d] &2456597.8330&0.0005\\
Inclination& $i$\, [deg]        & \multicolumn {2}{c}{50-70}& 3\\ 

\hline
\end{tabular}
\begin{minipage}{\textwidth}
\centering
$^1$~\citet{2021A&A...649A...1G}; 
$^2$~\citet{2021AJ....161..147B};
$^3$~this paper;
$^4$~Not taking account of extinction

\end{minipage}
\end{table*}

\subsection{Photometry}
\begin{figure*}
\includegraphics[width=0.95\textwidth]{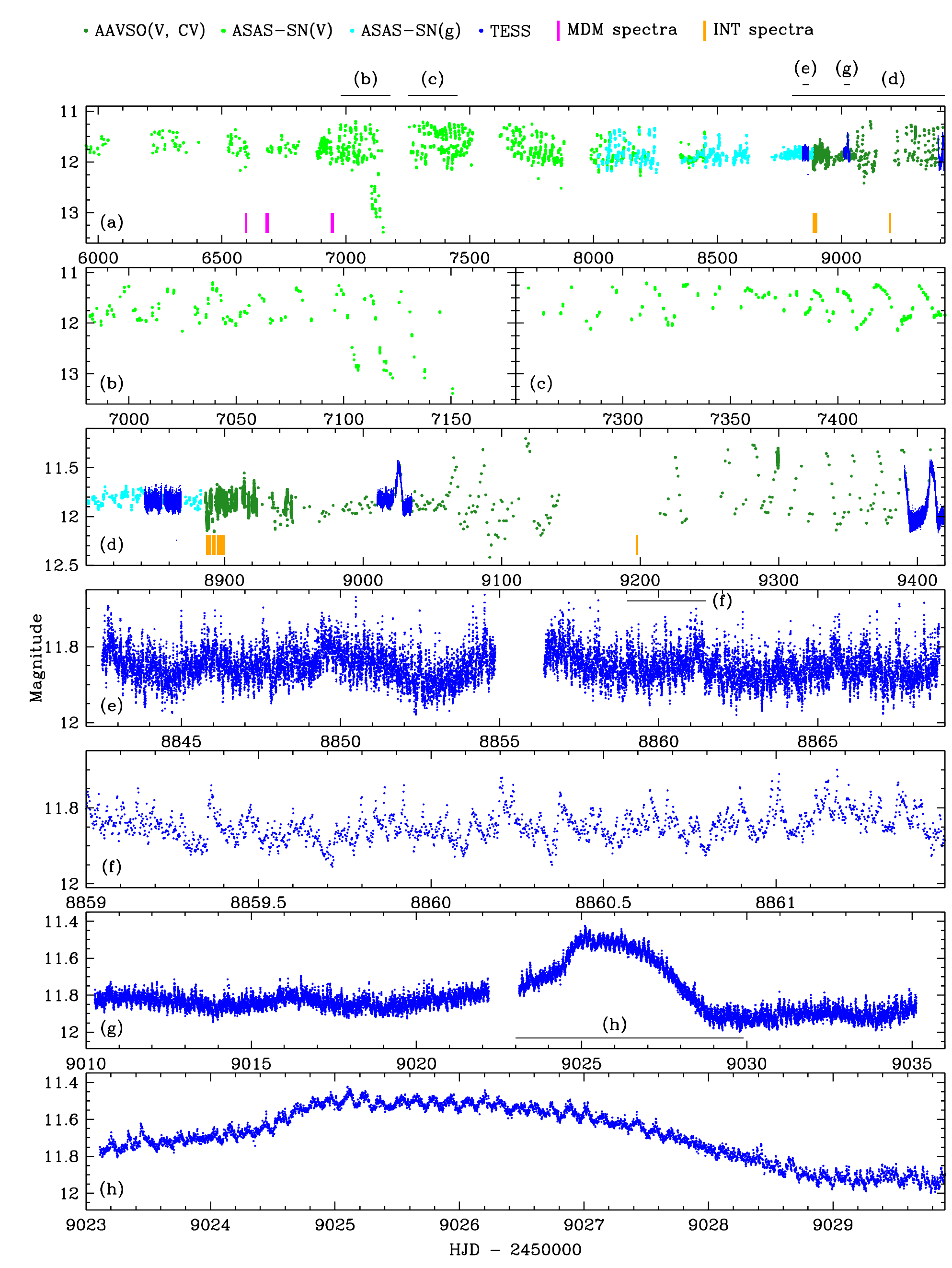}
\caption{Ground and space-based photometry of ASAS\,J0714+7004. The horizontal bars above the top panel, and within panels (e) and (g), indicate the date range of the zoom-in views. Note the change in the scale of the magnitude axis in the different panels. The system exhibits quasi-periodic brightenings most of the time, but experienced a fainter state (a,b) as well as a phase of relative inactivity (a,e,f). The high-cadence \textit{TESS} data reveals short-term flickering (e,f) and low-amplitude orbital modulations at the orbital period (g,h).} 
\label{fig:photometry}
\end{figure*}

The long-term light curve of ASAS\,J0714+7004 obtained by ASAS-SN \citep{2014ApJ...788...48S} displays quasi-periodic brightness variations on time scales of $\simeq20$\,d with an amplitude of $\simeq0.5$\,mag (panels (a), (b) and (c) in Fig.\,\ref{fig:photometry}). In order to investigate the nature of these variations, we began intensive photometric monitoring of the system using small-aperture telescopes throughout February and March 2020. However, during this campaign, ASAS\,J0714+7004 did not display much variability, remaining at an average magnitude of $V\simeq11.85\pm0.12$\,mag, although the quasi-periodic brightening resumed again in June 2020 (panel (d) in Fig.\,\ref{fig:photometry}). 

A peculiar event captured by the ASAS-SN data was a drop in the minimum brightness in between the quasi-periodic brightening in April 2015, whereas the maximum brightness remained constant (panel (b) in Fig.\,\ref{fig:photometry}). This ``low state'' lasted at least two months; unfortunately the ASAS-SN observations were stopped due to ASAS\,J0714+7004 moving into day time. When the observations resumed in August 2015, the system had returned to the normal state. 

The \textit{TESS} satellite \citep{2015JATIS...1a4003R} observed ASAS\,J0714+7004 in short-cadence (2\,min exposures) during Sector~20 (2019 December 24 to 2020 January 21) and Sector~26 (2020 June 8 to July 4) and in ultra-short cadence (20\,s exposures) during Sector~40 (2021 June 24 to July 23). The space-based photometry covers part of the inactive phase (panels (e) and (f) in Fig.\,\ref{fig:photometry}), the first brightening following this inactive phase (panels (g) and (h)), and two further brightenings (Fig.\,\ref{fig:photometry}, panel (d)). All the \textit{TESS} data sets reveal substantial periodic short-term variability.

\subsection{\textit{Swift}}
\label{sec:swift}
ASAS\,J0714+7004 was observed in January 2021 using the  Neil Gehrels  \textit{Swift} Observatory  \citep{2009MNRAS.397.1177E}. Data were obtained from both the ultraviolet optical telescope (UVOT) \citep{2005SSRv..120...95R} and X-ray telescope (XRT) \citep{2005SSRv..120..165B}. Ten observations were obtained over a 22\,d period with exposure times $\simeq10$\,min, see Table~\ref{tab:journal_Swift_observations}. Six of the ultraviolet observations used the $U$ filter (mean wavelength 3492\,\AA\ and width 866\,\AA) and four used the $M2$ filter (mean wavelength 2272.21\AA\ and width 533.85\,\AA). The XRT covers the range $0.2-10$\,keV and has two modes~--~Windowed Timing (WT) with continuous readout (and only one-dimensional imaging) and Photon Counting (PC) where the exposure length is 2.5\,s. We have only used PC data. The XRT data were processed using the automated pipeline to produce an average spectrum and a light curve. The UVOT data were saturated in both $U$ and $M2$ filters and therefore a modified approach \citep{2013MNRAS.436.1684P} was used to obtain photometry by measuring the ``streak'' as the CCD is read out.

\begin{table}
\caption [.] {\label{tab:journal_mdmobservations}  
Journal of MDM observations.
 }
\begin{tabular}{|l|l|l|l|l|l|}
\hline
UT (start) & Tel. & HA start & HA end & Exp(s) & $N_{\rm exp}$ \\
\hline
2013-11-01 09:39 & M & $-$02:19 & +00:55 &  600 &  3 \\
2014-01-22 11:05 & H & +04:30 & +05:04 &  480 &  4 \\
2014-01-26 11:06 & H & +04:47 & +05:05 &  240 &  4 \\
2014-01-27 06:22 & H & +00:06 & +00:50 &  480 &  4 \\
2014-10-12 12:37 & H & $-$00:40 & $-$00:32 &  480 &  1 \\
2014-10-13 12:49 & H & $-$00:24 & $-$00:19 &  300 &  1 \\
2014-10-14 12:55 & H & $-$00:15 & $-$00:04 &  300 &  2 \\
2014-10-15 12:10 & H & $-$00:56 & $-$00:06 &  300 &  9 \\
2014-10-16 07:17 & H & $-$05:46 & $-$00:03 &  240 & 17 \\
\hline
\end{tabular}
Telescopes are: M - McGraw-Hill 1.3 m; H - Hiltner 2.4 m.  Hour angles
are in hours and minutes. 
\end{table} 

\begin{table}
\caption [.] {\label{tab:journal_observations}  
The journal of INT observations for the February and December proposals.
 }
\begin{tabular}{|l|l|l|l|}
\hline
Date & JD (start) &N (observations)&Duration\\
& & &(minutes)\\ \hline
2020/02/08 & 2458887.583423 & 52 & 117.61 \\
2020/02/09 & 2458888.530952 & 52 & 124.56 \\
2020/02/10 & 2458889.465164 & 54 & 122.25 \\
2020/02/12 & 2458891.600877 & 2 & 4.20 \\
2020/02/13 & 2458892.455516 & 49 & 113.99 \\
2020/02/13 & 2458892.543982 & 44 & 117.47 \\
2020/02/16 & 2458895.603163 & 1 & 2.00 \\
2020/02/17 & 2458896.607681 & 2 & 4.20 \\
2020/02/18 & 2458897.602185 & 2 & 4.20 \\
2020/02/19 & 2458898.575209 & 2 & 4.20 \\
2020/02/20 & 2458899.526425 & 2 & 4.20 \\
2020/02/20 & 2458899.581305 & 2 & 4.20 \\
2020/12/13 & 2459197.430520 & 195 & 522.0 \\ \hline
\end{tabular}
\end{table}

\begin{table}
\caption [.] {\label{tab:journal_Swift_observations}  
The journal of \textit{Swift} observations. The central wavelengths of the $U$-band and $M2$-filter are 3492.67\,\AA\ and 2272.71\,\AA, respectively.}
\begin{tabular}{|l|l|l|l|l}
\hline
Date (UTC) & UVOT & XRT PC  & UV& UV Mag  \\
&exposure (s)&exposure (s)&Filter&(Vega)\\ \hline
2020-12-29 23:28:31 & 76.2 & 70.0 & $U$  &10.947(129)\\
2021-01-02 17:56:34 & 702.8 & 699.2 & $U$ &11.040(45\\
2021-01-04 00:13:36 & 583.5 & 574.7 & $M2$ &10.510(39) \\
2021-01-06 08:10:35 & 557.9 & 552.3 & $U$ &10.785(43)\\
2021-01-08 09:25:38 & 374.1  & 684.5 & $M2$ &9.828(32) \\
&  316.2 & & $M2$&9.680(32) \\           
2021-01-10 11:08:21 & 412.8 & 402.8 & $U$ & 10.409(38)\\
2021-01-14 08:56:35 & 425.2 & 423.7 & $U$ \\
2021-01-16 07:05:34 & 747.3 & 782.5 & $M2$ &10.627(38) \\
2021-01-18 08:32:35 & 564.5 & 557.5 & $U$ & 10.897(44)\\
2021-01-20 06:27:34 & 544.8 & 540.0 & $M2$ & 10.430(38)\\ \hline
\end{tabular}
\end{table}

\begin{figure} 
\includegraphics[width=\columnwidth]{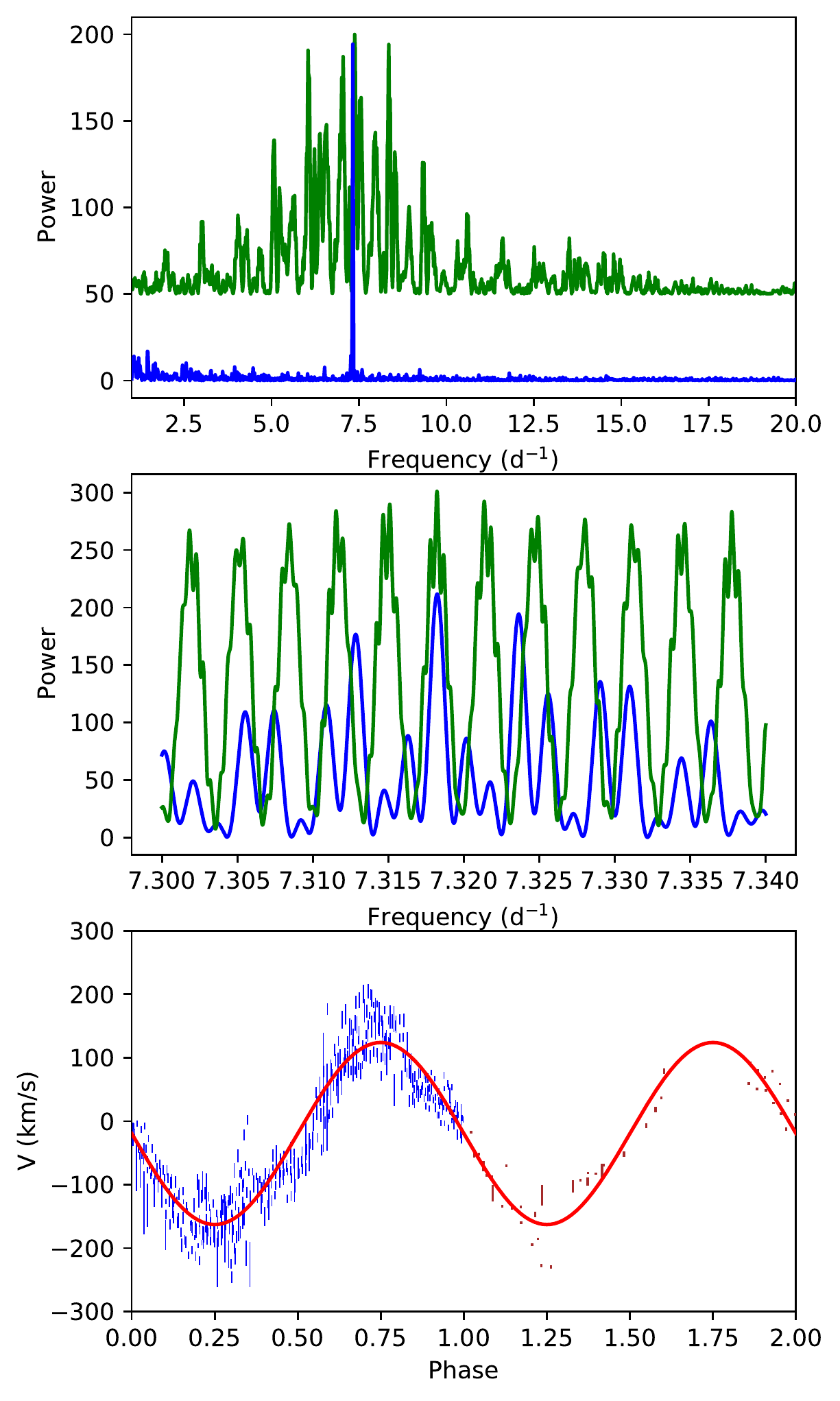}
\caption{\label{fig:Molly core}  
Top panel: Power spectra of the time-resolved radial velocity variations measured from the $H_{\alpha}$ line (offset in green) and the Sector~20,  26 and 40 \textit{TESS} light curves (blue). There is a clear peak at $\sim 3.28$\,h. Central panel: Detail from the top plots.  Bottom panel: The radial velocities of the INT data (blue) and the MDM data (brown) phase-folded using the adopted period of 3.2794894\,h, superimposed by a sine-fit to the data.}
\end{figure}

\section{Analysis} \label{sec:analysis}
The morphology of the spectroscopic and photometric data argues against a dwarf nova nature of ASAS\,J0714+7004. Although, at a first glance, the quasi-regular brightenings may seem unusual (see the discussion in Sect.\,\ref{sec:quasiperiodic}), we conclude that this system is best classified as a nova-like variable.

\subsection{Orbital period}
The primary motivation for obtaining the fast cadence INT observations was to measure the evolution of radial velocities, which we carried out using the \textsc{molly} software package \citep{2019ascl.soft07012M}. We also used \textsc{molly} to process the MDM spectra. We normalised the continuum of the spectra using a straight line fit, and then determined the central wavelength of the H$\alpha$ emission line by convolving each spectrum with a Gaussian template with a standard deviation of $\sigma=300\,\kms$. A Fourier analysis \citep{1975Ap&SS..36..137D} of the resulting INT and MDM radial velocity time-series was then used to compute a  periodogram (Fig.\,\ref{fig:Molly core} green line in the top panel). The highest peak in the periodogram identifies the orbital period as $P_\mathrm{orb}\simeq 3.28\,$h, albeit the periodogram is plagued by a substantial alias structure due to the sparse sampling of the radial velocities (Fig.\,\ref{fig:Molly core} middle panel).

The Fourier transform of the Sector~20,26 and 40  \textit{TESS} photometry results in a much less aliased periodogram (blue in Fig.\,\ref{fig:Molly core}) which, however, also contains some alias structure resulting from the gaps in the middle of the \textit{TESS} observations (panel (d) in Fig.\,\ref{fig:photometry}). We note that our time-series analysis of the \textit{TESS} data did not reveal any other periodic signal such as that seen in the nova-like V341\,Ara (which was interpreted as a beat frequency between the orbital period and the retrograde precession of the accretion disc in that system (Fig.\,8 in \citealt{2021MNRAS.501.1951C}).

The combined radial velocity and photometric periodograms overlap only for one unique period, which we therefore identify as the true orbital period of ASAS\,J0714+7004 (middle panel in Fig.\,\ref{fig:Molly core}).  In order to improve the precision of the orbital period, we performed a sinusoidal fit to the INT and MDM radial velocity data, which spans a much longer temporal baseline, with five distinct epochs, than the \textit{TESS} photometry, and hence provides a more accurate value of the period. We found $P=3.2794894(8)$\,h and an epoch of $\mathrm{HJD (UTC)}= 2456597.8330(5)$. We define the epoch as being the blue to red crossing of the  radial velocity. The velocity amplitude of the fit is $143.5(2.2)\,\kms$ The phase-folded data are shown in (Fig. \ref{fig:Molly core} bottom  panel).  A trailed spectrum of the December INT data is shown in Fig.\,\ref{fig:trailed_spectrum}. 

\begin{figure} 
\includegraphics[width=\columnwidth]{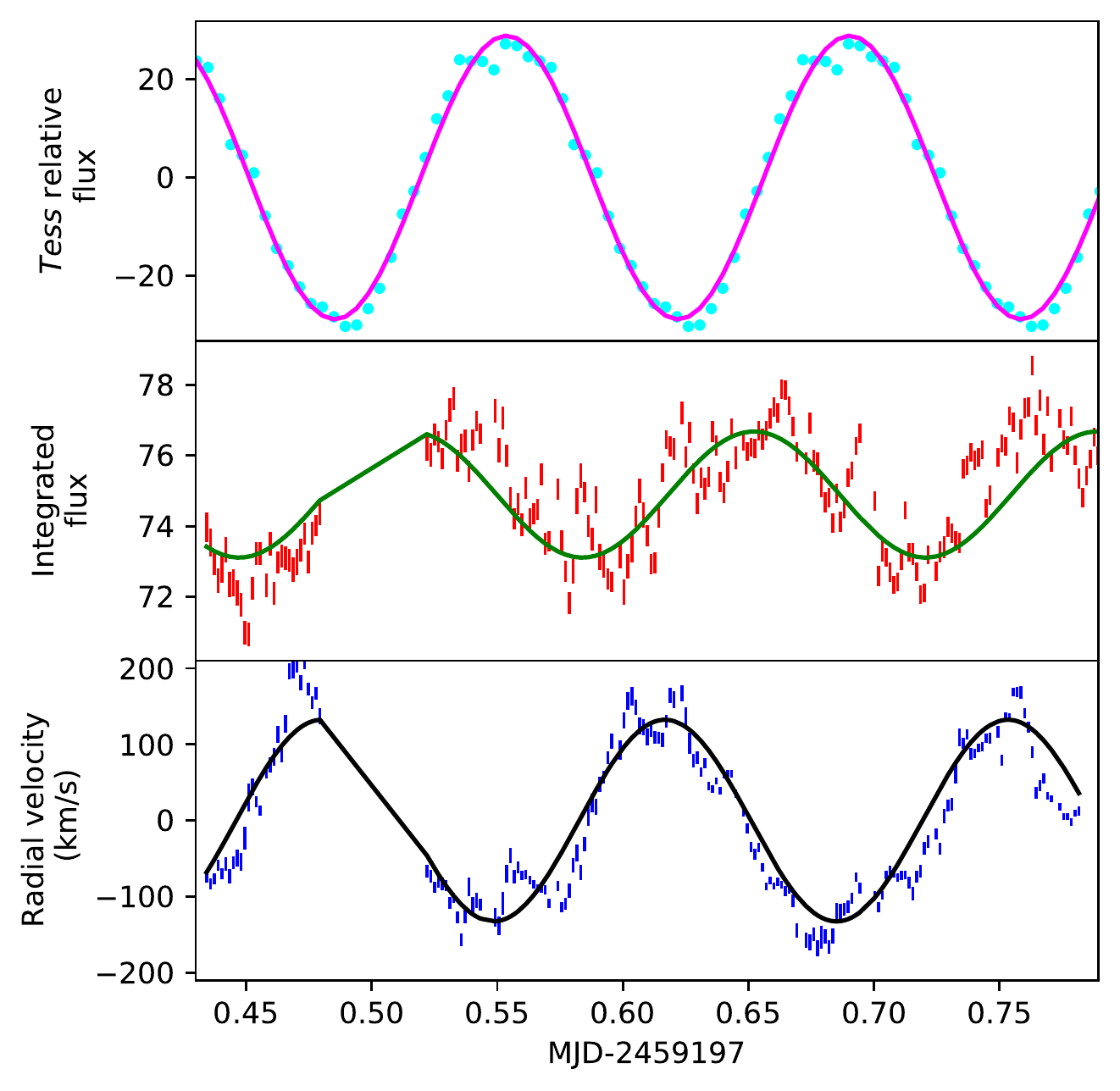}
\caption{\label{fig:integrated flux}  
The radial velocity of the H$\alpha$ core (bottom panel) and the integrated flux from the H$\alpha$ line (centre panel) for the December INT data together with the \textit{TESS} photometry binned and folded at the orbital period and projected onto the December timeline.  Each dataset was fitted with a sinusoid at the orbital period. The integrated flux is offset by 0.241 cycles from the radial velocities whilst the \textit{Tess} photometry is offset by 0.5 cycles.}
\end{figure}

The integrated flux from the H$\alpha$ line measured from the December INT data appears to lag the radial velocity variation by $\simeq90\degr$ (see Fig.\,\ref{fig:integrated flux}). Assuming that the  H$\alpha$ emission originates in the disc, then the radial velocity variation will track the movement of the disc (and hence the white dwarf at its centre) around the centre of gravity of the system. The peak of the integrated flux therefore occurs at the superior conjunction when the white dwarf is at its furthest point and the donor is at its closest. Thus, the relative phasing of the integrated line fluxes and radial velocities argues against the emission originating on the irradiated donor star.

\subsection{The mass transfer rate}
\label{sec:sed_fit}
A key parameter of CVs is their mass transfer rate, $\dot M$. This parameter is difficult to determine in dwarf novae, where the disc effectively buffers the mass lost from the companion \citep{2000A&A...361..952H}, and cyclically flushes it onto the white dwarf. However nova-likes such as ASAS\,J0714+7004 can be in a quasi-steady-state, facilitating the measurement of $\dot M$. Here, we estimate $\dot M$ in ASAS\,J0714+7004  by modelling its spectral energy distribution (SED). 

We have assembled the overall SED of ASAS\,J0714+7004 making use of available X-ray to infrared photometry (Pan-STARRS, \citealt{2016arXiv161205560C}; 2MASS, \citealt{2003yCat.2246....0C}; \textit{WISE}, \citealt{2014yCat.2328....0C}; \textit{Gaia},  \citealt{2020yCat.1350....0G}; ATLAS, \citealt{2018ApJ...867..105T}; APASS, \citealt{2015AAS...22533616H} and \textit{GALEX}, \citealt{2007ApJS..173..682M}) and of our \textit{Swift} observations (Sect.\,\ref{sec:swift}). 

\begin{figure*} 
\includegraphics[width=\textwidth]{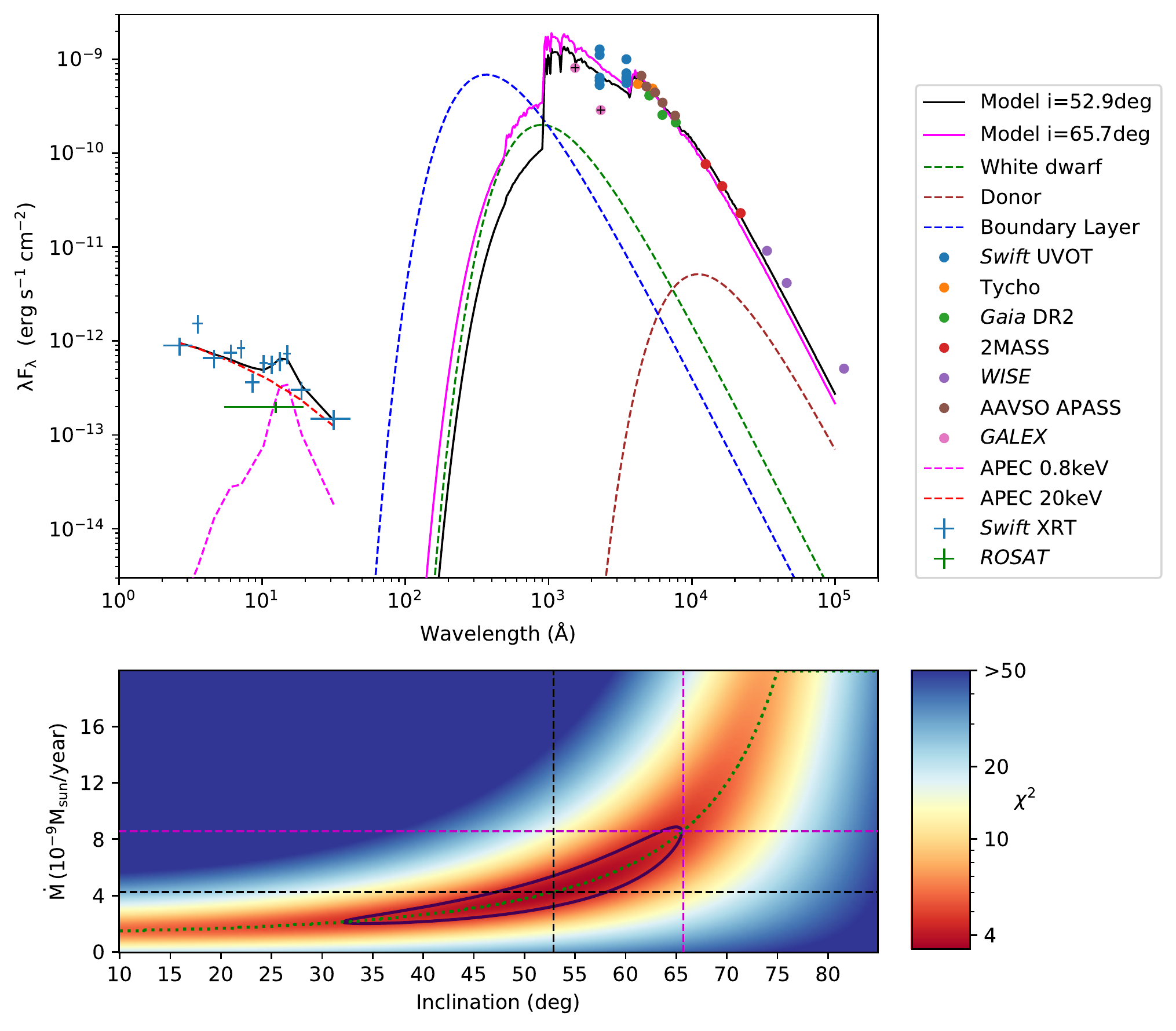}
\caption{\label{fig:SED}Multi-wavelength observations of ASAS\,J0714+7004 (see text for details). 
Top panel: Spectral energy density as a function of wavelength.   The disc dominates the SED in the optical; dashed lines show the contribution of the other components. The boundary layer contribution is only notionally constrained by the observations. The black solid line shows the optimum fit whilst the magenta line plots a fit constrained by the estimate of inclination  from Sect.\,\ref{sec:inc}. It is noticeable that the fits consistently undershoot the infrared observations. Such an infrared excess has been observed in other nova-likes \citep[e.g.][]{2014ApJ...786...68H}, indicating that the adopted model is not accounting for all components in the system. The excess may be due to a dusty circumstellar disc \citep{2007sptz.prop..299H,2009ApJ...693..236H}. 
Bottom panel: A heat map of the chi-squared of our model fit to the observed SED as a function of the two free parameters, $i$ and $\dot{M}$. 
The green dotted line traces the degeneracy between the orbital inclination and the mass transfer rate, illustrating that the two parameters cannot be determined independently. The black line indicates the best fit to the SED, and the magenta line  shows the estimate from Sect.\,\ref{sec:inc}. The solid purple line traces the  $1\sigma$ contour.  } 
\end{figure*}

The SED of ASAS\,J0714+7004 (Fig.\,\ref{fig:SED}) presents a steep blue slope from the infrared into the ultraviolet, with a significant drop in the spectral energy density in the X-ray regime. The system is sufficiently bright that the \textit{GALEX}  observations suffer from saturation effects \citep{2007ApJS..173..682M}~--~the \textit{Swift} UVOT photometry confirms that the \textit{GALEX} photometry is not reliable, and therefore it was not included in the fit.

The optical magnitude of ASAS\,J0714+7004 is $\simeq12$\,mag and this can also be expected to saturate the Pan-STARRS data (saturation occurs at $12-14$\,mag dependent upon seeing \citealt{2013ApJS..205...20M}) and the ATLAS data \citep{2021yCat..18670105T}.   The saturated points were therefore ignored when obtaining the best fit and are not shown (except for \textit{GALEX}) in Fig.\,\ref{fig:SED}. 

In order to model the SED of ASAS\,J0714+7004, we have to account for the flux emitted by the individual components of the system, i.e. the disc, the white dwarf, the donor star and potentially also a hot boundary layer close to the white dwarf where up to half of the gravitational potential energy is released. The detailed structure, and even the presence of a boundary layer in CVs has been subject to decades of discussion \citep{1982ApJ...262L..53F, 1996ApJ...469..841L, 2013A&A...560A..56H, 2019A&A...628A.121N}, and, on the observational side, is severely limited by the fact that interstellar absorption severely hinders access to the extreme ultraviolet (EUV) emission of CVs\footnote{The boundary layer is expected to have temperatures of a few $10^5$\,K, and hence its emission will peak in the Extreme Ultraviolet. This wavelength region is heavily absorbed by interstellar neutral hydrogen, and hence only a handful of nearby CVs have adequate observations, which led to conflicting  results:  \citet{1996ApJ...469..841L} observed U\,Gem during outburst and concluded that the peak spectrum resembled a blackbody with $\Teff \simeq 140\,000$\,K. Conversely a joint \textit{ROSAT} and \textit{EUVE} data analysis of the nova-like IX\,Vel  showed no blackbody emission \citep{1995A&A...300..808V}.}. In the case of ASAS\,J0714+7004, we note that whereas a putative boundary layer may contribute in the EUV, it will make only a minimal contribution when fitting the ultraviolet to infrared emission of the system (see Fig. \ref{fig:SED}). Here, we model the SED of ASAS\,J0714+7004 by the sum of the disc, the white dwarf and donor and the boundary layer. In nova-likes, the accretion disc is expected to be the primary source of optical and ultraviolet flux, in contrast the two stellar components have only relatively small contributions.

We model the disc as a set of 1000 concentric annuli of constant width between the inner and outer radii of the disc, $R_\mathrm{disc,i}$ and $R_\mathrm{disc,o}$, respectively. To determine the temperature $T(r)$ of each annulus we assume that the flux per unit surface area at radius $r$ in the disc is given by (see chapter~5 of \citealt{1992apa..book.....F}):
\begin{equation}
F = \frac {3GM_1 \dot{M}}{8\pi  r^{3} } \left( 1 -\left( \frac{R_1}{r} \right) ^ {1/2} \right)   \label{eq:scale_heights}
\end{equation}
where $M_1$ and $R_1$ are the mass and radius, respectively, of the white dwarf and $\dot{M}$ is the rate of mass transfer. Given that the integrated flux is a function of the temperature, \begin{equation}
F = \sigma T^{4}
\end{equation}
we obtain
\begin{equation}
T = \left(\frac {3GM_1 \dot{M}}{8\pi \sigma r^{3} } \left( 1 -\left( \frac{R_1}{r} \right) ^ {1/2} \right)   \right)^{1/4} \label{eq:model_temp}
\end{equation}

\noindent
We then used the BT-Settl (AGSS2009) spectral models  \citep{2012RSPTA.370.2765A} to represent the spectrum for each annulus based on the temperature from Eq.\,\ref{eq:model_temp}, $\log g=4.0$ and solar metallicity. The shortcomings of using synthetic stellar spectra to model the emission of an accretion disc have been discussed at length,  \citep{1988ApJ...335..394W} however they are better than the alternative which is to use a blackbody spectrum.

The white dwarf temperature was treated as a function of $\dot{M}$ using Eq. 2 in \citep{2009ApJ...693.1007T}:
\begin{equation}
T_1 = 17\,000\mathrm{[K]}\left(\frac { \langle \dot{M}\,[\Msy] \rangle}{10^{-10}}
\right)^{\frac{1}{4}} \left(\frac{M_1\,[\Msun]}{0.9} \right)
 \label{eq:T_wd}
\end{equation}

For the donor star, we adopted a mass, radius, and effective temperature based upon the evolutionary track from Table\,2 by \citet{2011ApJS..194...28K}. The emission of the white dwarf and the donor were modelled by two blackbodies of the corresponding temperatures and radii.  

The fit shown in Fig. \ref{fig:SED} was obtained by fixing $M_1=0.83\,\Msun$  (the average value for CVs from \citealt{2011A&A...536A..42Z}) and $R_1=0.0122\,\Rsun$ (using the mass/radius relationships from \citealt{2020ApJ...901...93B} available at \footnote{\url{https://www.astro.umontreal.ca/~bergeron/CoolingModels/}}).

We also followed \citet{2017ApJ...846...52G} and assumed that the inner edge of the accretion disc is located one $R_1$ above the white dwarf surface. We followed \citet{1996MNRAS.279..219H} and assumed  that the outer radius of the disc (due to tidal truncation) is $0.7 \times R_{\textrm{L1}}$ where $R_{\textrm{L1}}$ is the distance from the centre of the white dwarf to the inner Lagrange point given by:
\begin {equation}
 R_{\textrm{L1}}= a \left(1.0015+q^{0.4056}\right)^{-1}
\end {equation}
The inclination of the accretion disc and ${\dot{M}}$ were treated as free parameters.

We assumed extinction based on the galactic formulae in \citep{ 1983MNRAS.203..301H}.  We then fitted the data using the system parameters in table\,\ref{tab:model} and a $320\times 600$ linearly spaced grid of values of $\dot{M}$ from $0.5-20\times10^{-9}\,\Msy$  , and inclination $10-85\degr$  to produce the heat map in Fig.\,\ref{fig:SED}. In order to account for the fact that the object is intrinsically slightly variable, we adopted for the fit either the published uncertainties on the photometric data, or a value of 10~per cent, whichever was larger.

To demonstrate the contribution of a potential boundary layer we also modelled the effect of adding a circular band with $H= 0.1 R_1$, $R=R_1$ and $T_\textrm{bl,eff}=100\,000$\,K. This does not significantly affect the fit as its emission in the optical and near-ultraviolet is far less than that of the disc (see Fig. \ref{fig:SED}).

\begin{table}
\caption{\label{tab:model} System parameters adopted for fitting the spectral energy distribution. }
\centering
\begin{tabular}{ l l c  } 

\hline
\multicolumn {2}{l}{Parameter} & 
\multicolumn {1}{
c}{Value}   \\
\hline

Mass (white dwarf)               & $M_{1}$\,[$\Msun$]           & 0.83 \\ 
Mass (donor)               & $M_{2}$\,[$\Msun$]           & 0.22 \\
Separation   & $a$\, [$\Rsun$]           & 1.11  \\ 
Radius (white dwarf)            & $R_{1}$\, [$\Rsun$]  &0.0122  \\ 
Radius (donor)                  & $R_{2}$\, [$\Rsun$]           & 0.30  \\ 
Inner radius (disc)          & $R_\mathrm{disc,i}$\, [$\Rsun$]           & 0.0244 \\ 
Outer radius (disc)    & $R_\mathrm{disc,o}$\, [$\Rsun$]           & 0.49  \\ 
 
Temperature (donor)& $T_\mathrm{2}$\, [K]        & 3324\\ 

\hline
\end{tabular}

\end{table}

There is considerable degeneracy between the angle of inclination and $\dot{M}$. This is evident from the dotted green line in Fig.\,\ref{fig:SED} showing the best fit for a given inclination. The overall best fit is found for $i=52.9\degr$ (see table \ref{tab:results}), consistent with the estimate calculated in section \ref{sec:inc} ($65.7\degr \substack{+8\\-13}$ ). The disc luminosity is better constrained, $3\pm1  \times 10^{33}\,\mathrm{erg\,s^{-1}}$ (see Table\,\ref{tab:results}) . 

The white dwarf temperature determined from the SED fit only mildly depends on the inclination (Table\,\ref{tab:results}) and is consistent with those found in other nova-likes and VY\,Scl stars from direct modelling of the white dwarf spectrum: DW\,UMa ($\simeq50\,000$\,K,  \citealt{2003ApJ...583..437A}), TT\, Ari ($\simeq39\,000$\,K, \citealt{1999A&A...347..178G}) and MV\,Lyr ($\simeq47\,000$\,K, \citealt{2004ApJ...604..346H}).
\begin{table}
\caption{\label{tab:results} Estimated values from fitting the spectral energy distribution. }
\centering
\begin{tabular}{ l l c c } 

\hline
\multicolumn {2}{l}{Parameter} & 
\multicolumn {1}{c}{i=52.9 $\deg$} &
\multicolumn {1}{c} {i=65.7 $\deg$}   \\
\hline

Mass transfer rate               & $\dot{M}$\,[$10^{-9}\Msy$]           & 4.75&8.13 \\ 
Temperature & $T_\mathrm{1}$\, [K]        & 41\,158&48777\\ 
(white dwarf)\\
Disc luminosity               & $L_{\mathrm{disc}}$\,[$\mathrm{10^{33} erg\,s^{-1}}$]           & 2.59&3.66 \\

\hline
\end{tabular}

\end{table}

\subsection{Orbital inclination}\label{sec:inc}
\begin{figure} 
\includegraphics[width=\columnwidth]{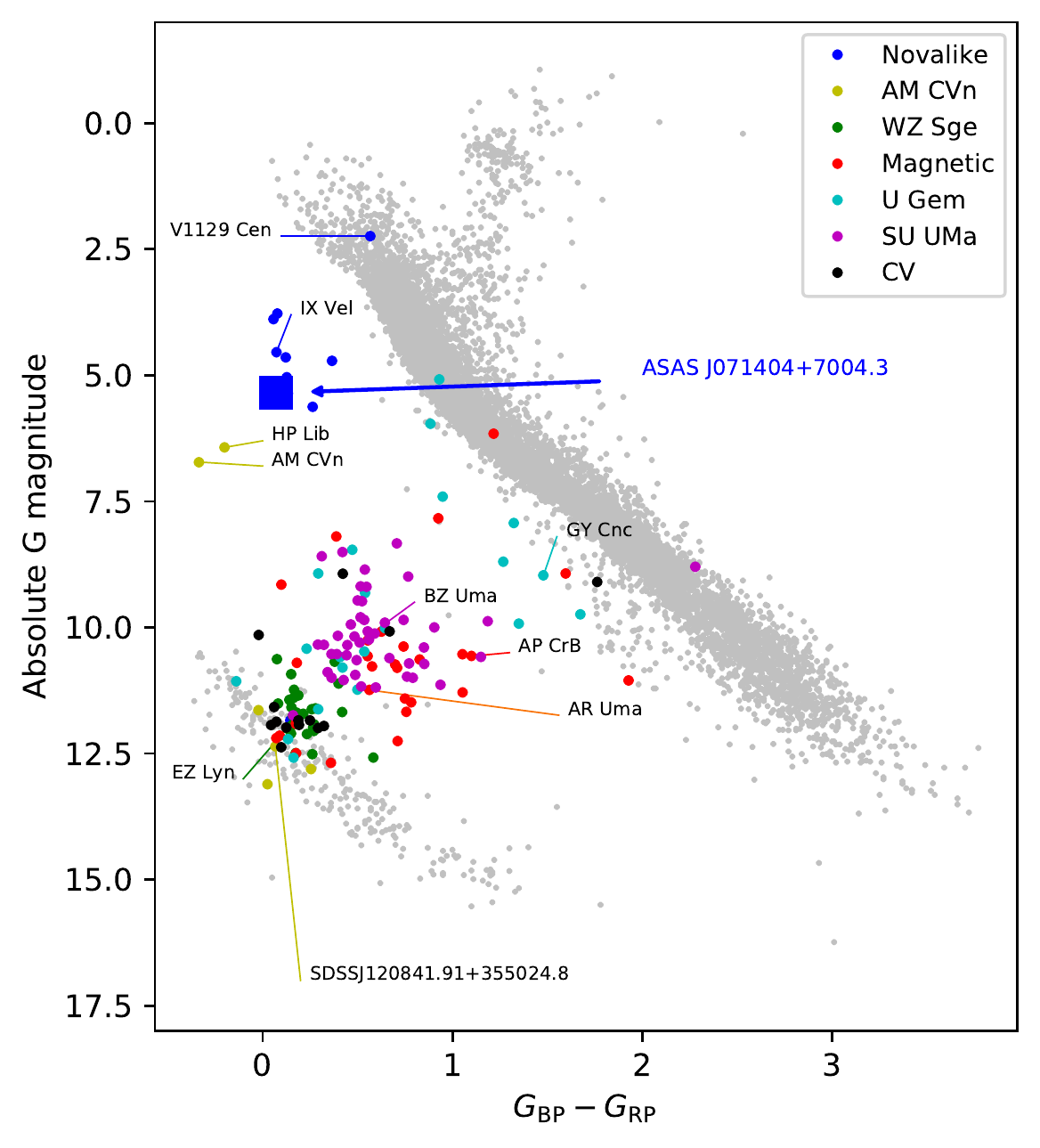}
\includegraphics[width=0.95\columnwidth,right]{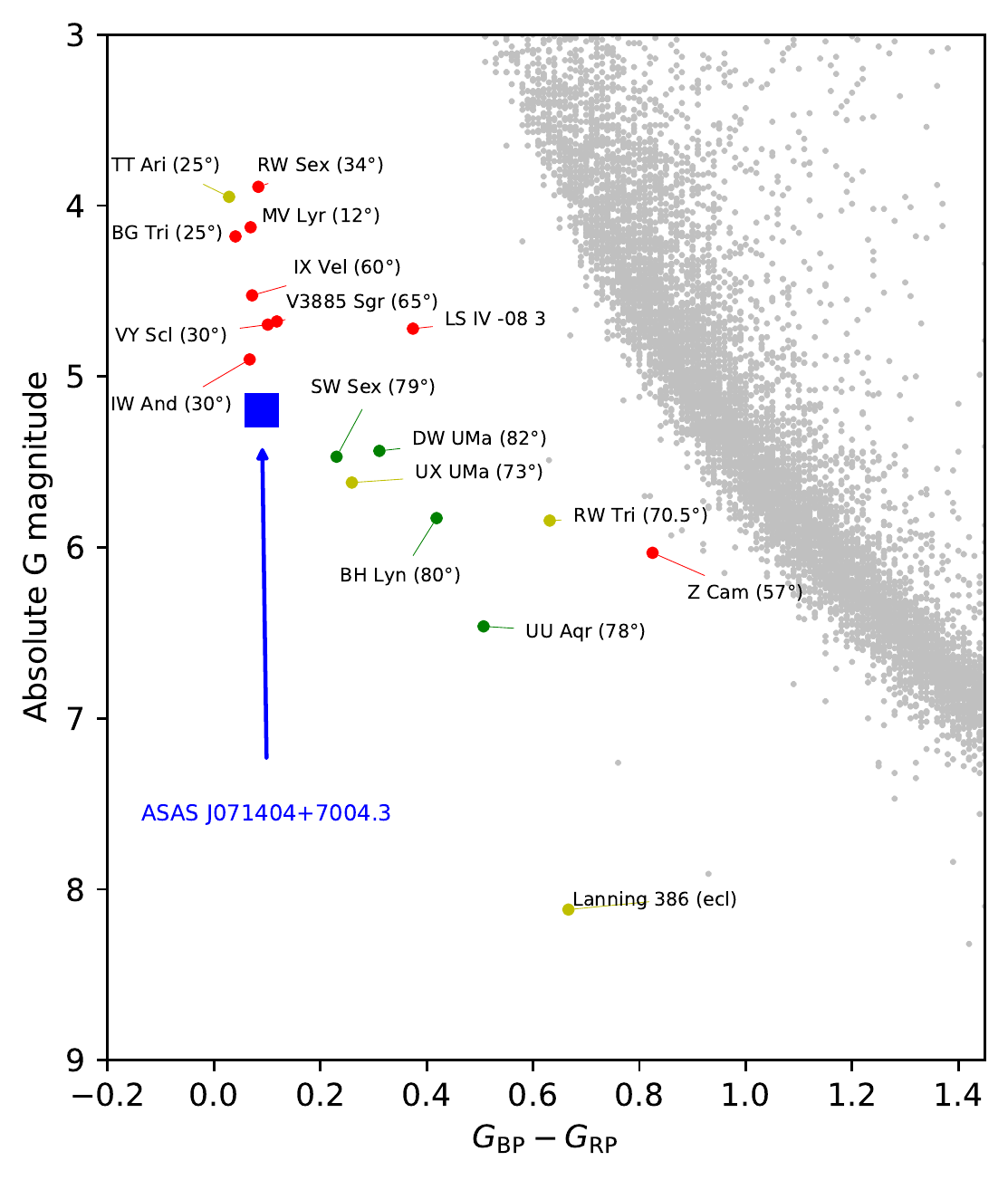}
\caption{\label{fig:HR_diagram}  
Top panel: The ``Gold Sample'' of known CVs within 300pc with reliable astrometry and astrometry from \citet{2021MNRAS.504.2420I}.  ASAS\,J0714+7004 is marked with a blue square. Bottom panel: Detail showing the area of the HRD occupied by a representative selection of well-studied nova-likes. Confirmed SW\,Sex stars are shown using green dots and candidate SW\,Sex stars using yellow. The data is from \textit{Gaia} EDR3 \citep{2020arXiv201201533G} without adjusting the absolute magnitudes for extinction. The inclination (where known) is shown in brackets (sources 
 'SW Sex', \citep{ 2001A&A...368..183G}
 'BH Lyn', \citep{ 1992MNRAS.258..225D}
 'DW UMa', \citep{ 2000A&A...364..573B}
 'Lanning 386', \citep{ 2008PASP..120..301B}
 'MV Lyr', \citep{ 2004ApJ...604..346H}
 'BG Tri', \citep{ 2021MNRAS.503.1431H}
 'RW Tri', \citep{ 2019AcA....69...79S}
 'UX UMa', \citep{ 2004RMxAC..20..270S}
 'TT Ari', \citep{ 2019MNRAS.489.2961B}
 'UU Aqr', \citep{ 1994ApJ...433..332B}
 'RW Sex', \citep{ 1992A&A...256..433B}
 'IW And', \citep{ 2013PASP..125.1421S}
 'VY Scl', \citep{ 2000ApJ...538..315M}
 'Z Cam', \citep{ 1981ApJ...251..201S}
 'V3885 Sgr', \citep{ 2002MNRAS.332..127H}
 'IX Vel', \citep{ 2002MNRAS.332..127H}). 
Lanning\,386 is an interesting outlier. It is  an eclipsing system  \citep{2017MNRAS.466.2202K, 2016AJ....152...27K} and more significantly it is most often, unusually for a nova-like, in its low state.}
\end{figure}

As discussed in Sect.\,\ref{sec:sed_fit}, the inclination of ASAS\,J0714+7004 is not tightly constrained from the SED fit. Here, we will follow an alternative approach to estimate the inclination based on the absolute magnitude of the system. Its location within the Hertzsprung-Russell diagram (HRD) places it firmly among the nova-likes (Fig.~\ref{fig:HR_diagram}), which are characterised by high mass-transfer rates that keep their accretion discs in the hot state. We therefore adopt the absolute magnitude-orbital period relation of  \citeauthor{1980AcA....30..127P} (\citeyear{1980AcA....30..127P}), see also (\citealt{1987MNRAS.227...23W} )
\begin{equation}
M_{\mathrm{}} = 5.7\pm0.13 - 0.287\pm0.024\times P_\mathrm{orb} 
 \label{eq:abs_mag}
\end{equation}
which is adjusted for inclination by:
\begin{equation}
\Delta M_{\mathrm{v}}=-2.5 \times \log \left(\left(1+\frac{3}{2} \cos i\right)\cos i\right)  \pm{0.4}
 \label{eq:delta_mag}
\end{equation}
Based on the \textit{Gaia} $G$-band data in Table \ref{tab:ephemera}, the mean absolute $G$-band magnitude of ASAS\,J0714+7004 is $5.20 \pm{0.016}$. Using Eq.\,\ref{eq:abs_mag} and the orbital period of 3.28\,h gives a predicted absolute magnitude of $4.76\substack{+0.21\\-0.16}$ and therefore $\Delta M_{\mathrm{V}}=0.44\pm0.21$. Solving  Eq.\,\ref{eq:delta_mag} for $i$ (and taking account of the error in the formula  of  $\pm{0.4}$)  yields an inclination of $65.7\substack{+8\\-13}\degr$. 
This estimate is also consistent with the SED fit, the strong Balmer emission lines and the fact that it is not eclipsing. It is also consistent with Figure\,1 in \citet{2018AJ....156..198H}. 

The \textit{Gaia} astrometry provides the opportunity to test the absolute magnitude-orbital period relation: most nova-likes, including ASAS\,J0714+7004, have orbital periods of $3-4$\,h, and hence their absolute magnitudes should primarily depend on their inclinations. Inspecting  Fig.~\ref{fig:HR_diagram} (lower panel), this is indeed the case, where the fainter systems have higher inclinations~--~in particular the green and yellow coded SW\,Sex systems are eclipsing and will therefore inevitably have large inclinations. We caution, however, that some care needs to be taken when considering cataclysmic variables in the \textit{Gaia} HRD, as their exact location is subject to the state(s) that the \textit{Gaia} sampled. As such, CVs may move by some amount within the HRD when comparing their DR2 and EDR3 data (see Appendix\,\ref{sec:appendix}).  

\subsection{Photometric variability}
\begin{figure}
    \centering
    \includegraphics[width=\columnwidth]{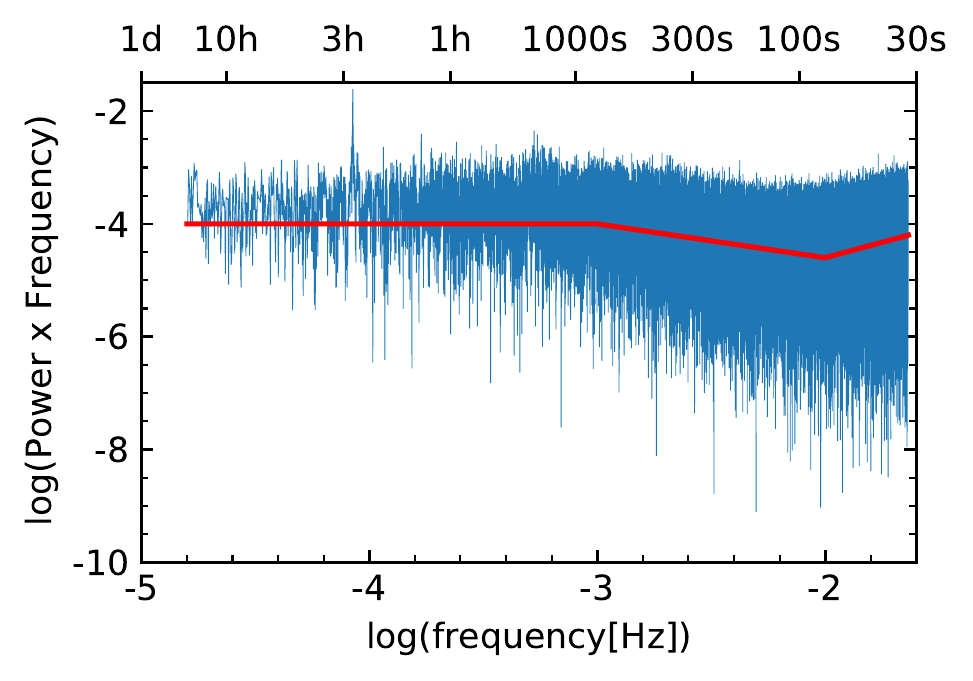}
    \caption{Power density spectrum of ASAS\,J0714+7004 computed from the Sector\,40 observations, which were obtained with at a 20\,s cadence. The orbital period of 3.28\,h is clearly detected. The slight peak at $\simeq$2000s does not appear in the data of the other sectors. It first appears with a period of 1883\,s on day 5 and varies between 1594s and 1931s before disappearing after day 11. It may be related to intermittent quasi-periodic oscillations. The overall shape of the broad-band noise can be fit with three simple power laws. The frequency break at $\approx 10^{-3}$\,Hz is typical of accretion-induced variability. The rise in power at $\approx 10^{-2}$ Hz is attributed to Poisson noise dominating the power spectrum.}
    \label{fig:tess_20s_power}
\end{figure}

The uninterrupted \textit{TESS} observations provide the opportunity to investigate the broad-band variability of ASAS\,J0714+7004 on timescales of minutes to days. The Fourier transform \citep{1975Ap&SS..36..137D} computed from the Sector~40 data that was obtained in the ultra-short 20\,s cadence mode (Fig.\,\ref{fig:tess_20s_power}) shows clearly the 3.28\,h orbital period, and a flat distribution ($-1$ powerlaw) in power density down to $\simeq6.3$\,mHz, after which the power law breaks to $-2$. A weak signal at 2000\,s was detected in the Sector\,40 data, but was absent in the other \textit{TESS} observations, and may have been related to intermittent quasi-periodic oscillations \citep{2002MNRAS.335...84W}. The power law slopes and break frequency are typical of those observed in other accreting white dwarfs which display strong broad-band variability components (see e.g. MV\,Lyr,  \citealt{scaringi12,scaringi14,scaringi15}). This phenomenology is generally explained by the so-called fluctuating accretion disc model, which was originally developed to explain the broad-band noise properties observed in X-ray binaries and active galactic nuclei (see \citealt{lyubarskii97,AU06,ID11,ID12,IV13}). In this model mass transfer variations through the disc couple multiplicatively as matter is moved inwards through the disc. The observed power spectrum is then the result of the faster variability driven by mass transfer fluctuations and emitted in the inner-disc regions \citep{scaringi15}. In reality the variability is generated across a range of disc radii, and is observed in a specific pass-band given an emissivity profile for the disc. A similar power spectral density observed in the novalike MV\,Lyr has been modelled by \citet{scaringi14} using the fluctuating accretion disc model. From the modelling, the power spectral break reveals that the variability is driven by a geometrically thick (and optically thin) disc component which extends from the white dwarf surface up to a few white dwarf radii \citep{scaringi14}. By analogy, the same disc component, as inferred from the break in the power spectrum shown in Fig.\,\ref{fig:tess_20s_power} may also be present in ASAS\,J0714+7004. This would also be consistent with the inference of the boundary layer discussed in Sect.\,\ref{sec:sed_fit}, although the geometrical extent inferred from the broad-band variability would suggest this component to be substantially extended.    

\subsection{X-ray emission}
ASAS\,J0714+7004 was observed by \textit{ROSAT} during its all sky survey \citep{1999A&A...349..389V}.  The \textit{ROSAT}  Position Sensitive Proportional Counter (PSPC) observation, covering 0.1--2.4\,keV, detected a count rate of $0.17\pm 0.023\,\mathrm{cts\,s^{-1}}$. We adopted the flux based on a fit of a stellar model to the PSPC data ($\mathrm{Flux2}=1.92\times10^{-15}\,\mathrm{W\,m^{-2}}$)  to illustrate the \textit{ROSAT} detection in Fig.\,\ref{fig:SED}.   The stellar model \citep{1999A&A...349..389V} is based on an empirical conversion between count-rates and fluxes following \citep{1995ApJ...450..392S} that was originally developed to obtain flux values for stars.

\begin{figure} 
\includegraphics[width=\columnwidth]{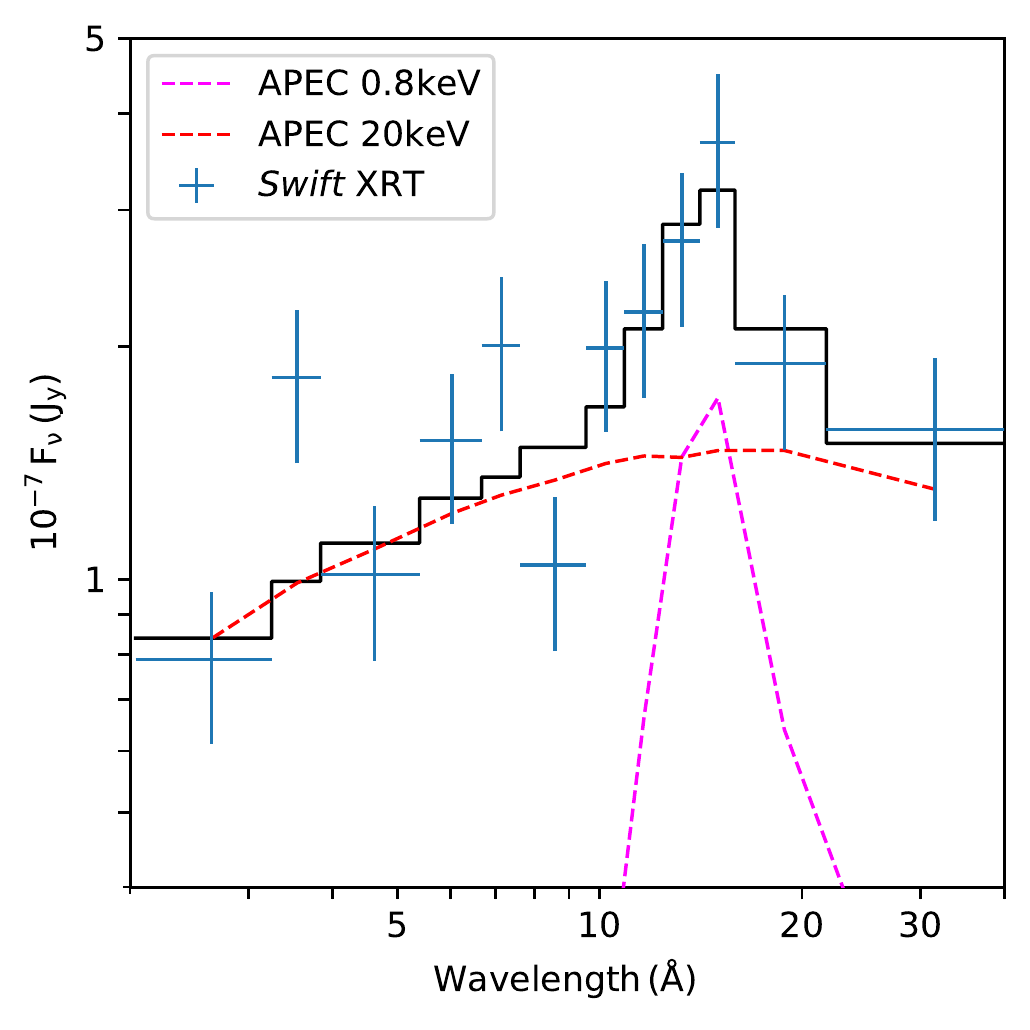}
\caption{\label{fig:x-ray spectrum}   Spectrum of the \textit{Swift} XRT observations (crosses). We assumed  interstellar absorption fixed at $N_\mathrm{H}=3.2 \times 10^{19}\,\mathrm{cm^{-2}}$ and solar abundances. The spectrum was then fitted using the sum of two APEC components with temperatures of $0.8 $ keV  and $20$ keV respectively, and the sum is shown as the grey line, binned to the observed energy channels. 
}
\end{figure}

Our \textit{Swift} XRT spectrum consists of a set of counts binned by energy (see Fig. \ref{fig:x-ray spectrum}) in the range 0.3 to 6\,keV. To obtain an estimate for the luminosity we fitted the data using \textsc{xspec}. We have used a model with two components~--~one with a temperature over 16\,keV and the other with a  temperature of $0.83\pm{0.15}$\,keV. This model was then used to extrapolate upwards into harder X-rays and downwards into the EUV (see Table \ref{tab:x-rayflux}). Absorption is based on assuming 
an interstellar hydrogen column density of $N_\mathrm{H}=3.2 \times 10^{19}\,\mathrm{cm^{-2}}$ ; in practice this has a minimal effect.

\begin{table}
\caption{\label{tab:x-rayflux} Flux and luminosity based on \textit{Swift} XRT observations.}
\begin{tabular}{|l|c|c|c|c|}
\hline
\multirow{2}{*}{} & \multicolumn{2}{c}{Range} & Flux                                                 & X-ray luminosity                     \\ \cline{2-3}
                  & keV         & \AA           & ($\mathrm{10^{-12} erg\,s^{-1}\,cm^{-2}}$) & ($\mathrm{10^{30} erg^{-1}}$) \\ \hline
Observed          & 0.3--6.0       & 2.1--41       & 1.67                                                 & 9.07                           \\
Extrapolated      & 0.3--10      & 1.2--41       & 2.33                                                 & 12.65                          \\ 
Extrapolated      & 0.1--10      & 1.2--124      & 2.57                                                 & 13.95                          \\
Extrapolated      & 0.01--10      & 1.2--1242     & 2.66                                                 & 14.44                          \\ \hline
\end{tabular}
\end{table}

The \textit{Swift} pipeline spectrum is shown in Fig.\,\ref{fig:x-ray spectrum} and qualitatively resembles those of BZ\,Cam and MV\,Lyr (Figures 2 and 3 in \citealt{2014ApJ...794...84B}). The X-ray luminosity of ASAS\,J0714+7004 of $1.17\pm{0.27} \times 10^{31}\,\mathrm{erg\,s^{-1}}$
is lower than that of some dwarf novae \citep{2010MNRAS.408.2298B}, although an order of magnitude greater than that of the average CV \citep{2013MNRAS.430.1994R}. The optically-thin spectrum and inferred temperatures are similar to other non-magnetic CVs \citep{2005MNRAS.357..626B,2002ApJ...566L..33M}.

We inspected the average count rates for each of our \textit{Swift} XRT observations , and we note that there is mild ($1\,\sigma$) evidence for variability across the 22\,d spanned by the data.

\begin{figure} 
\includegraphics[width=\columnwidth]{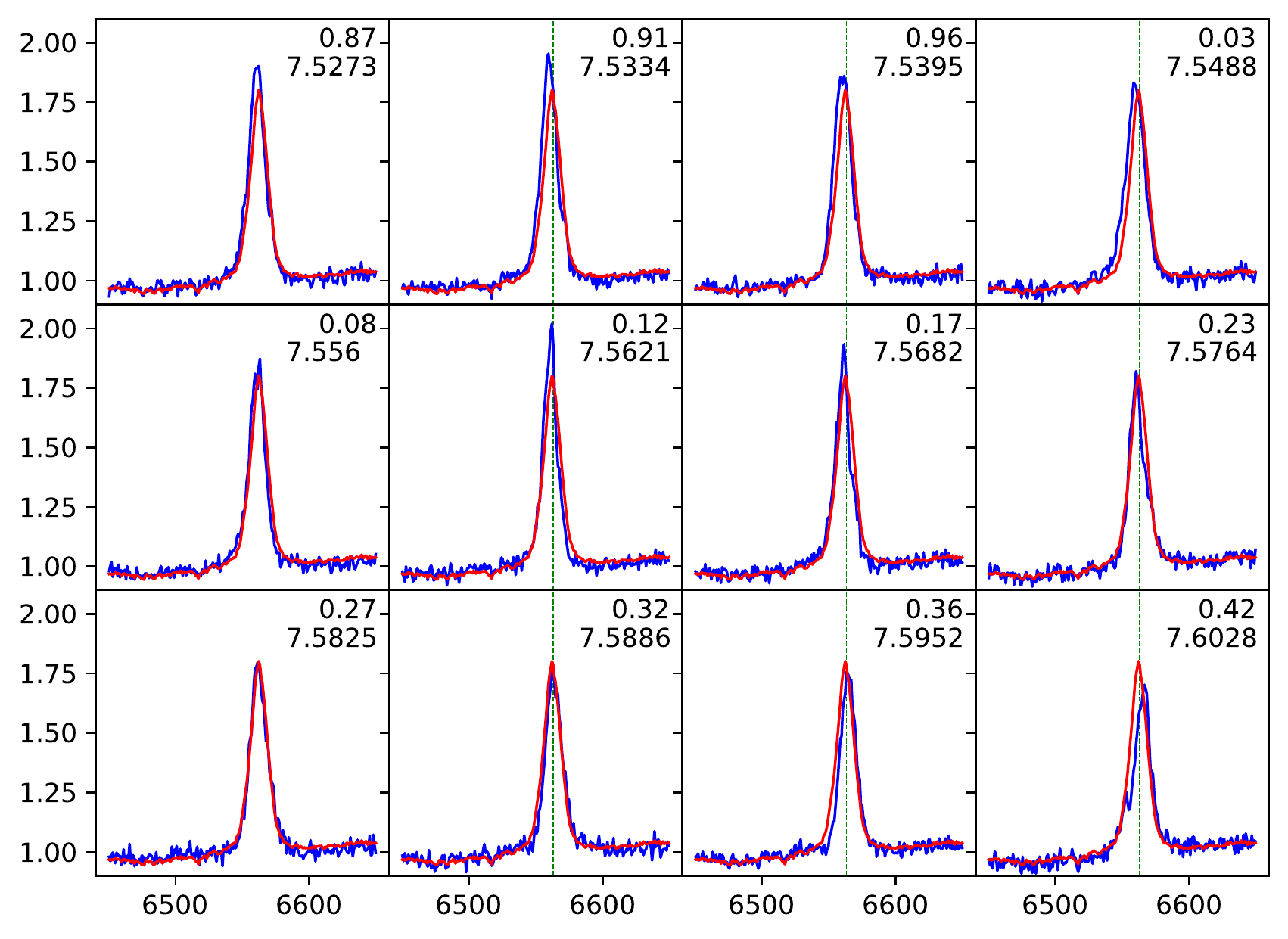}
\includegraphics[width=\columnwidth]{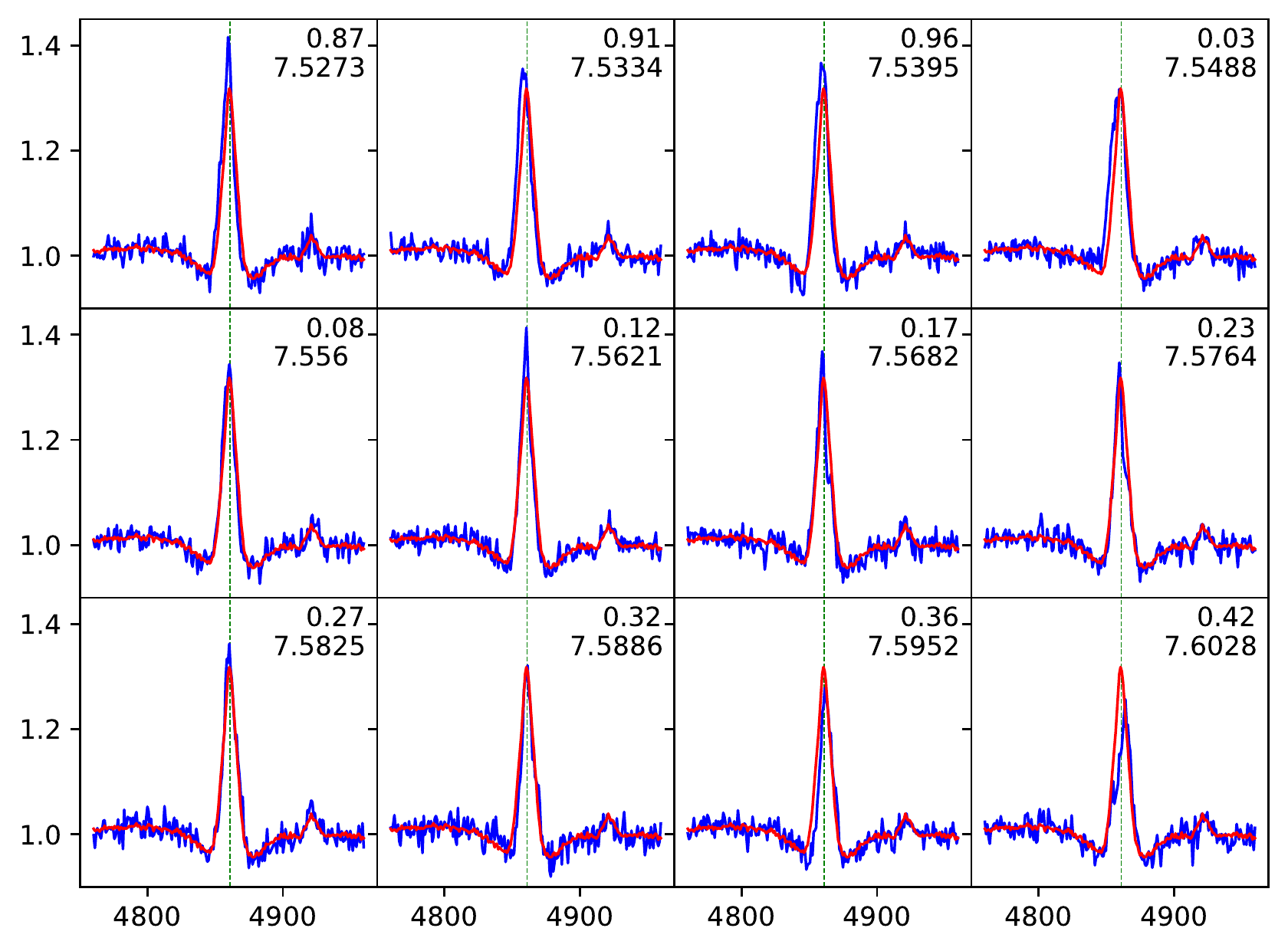}
\includegraphics[width=\columnwidth]{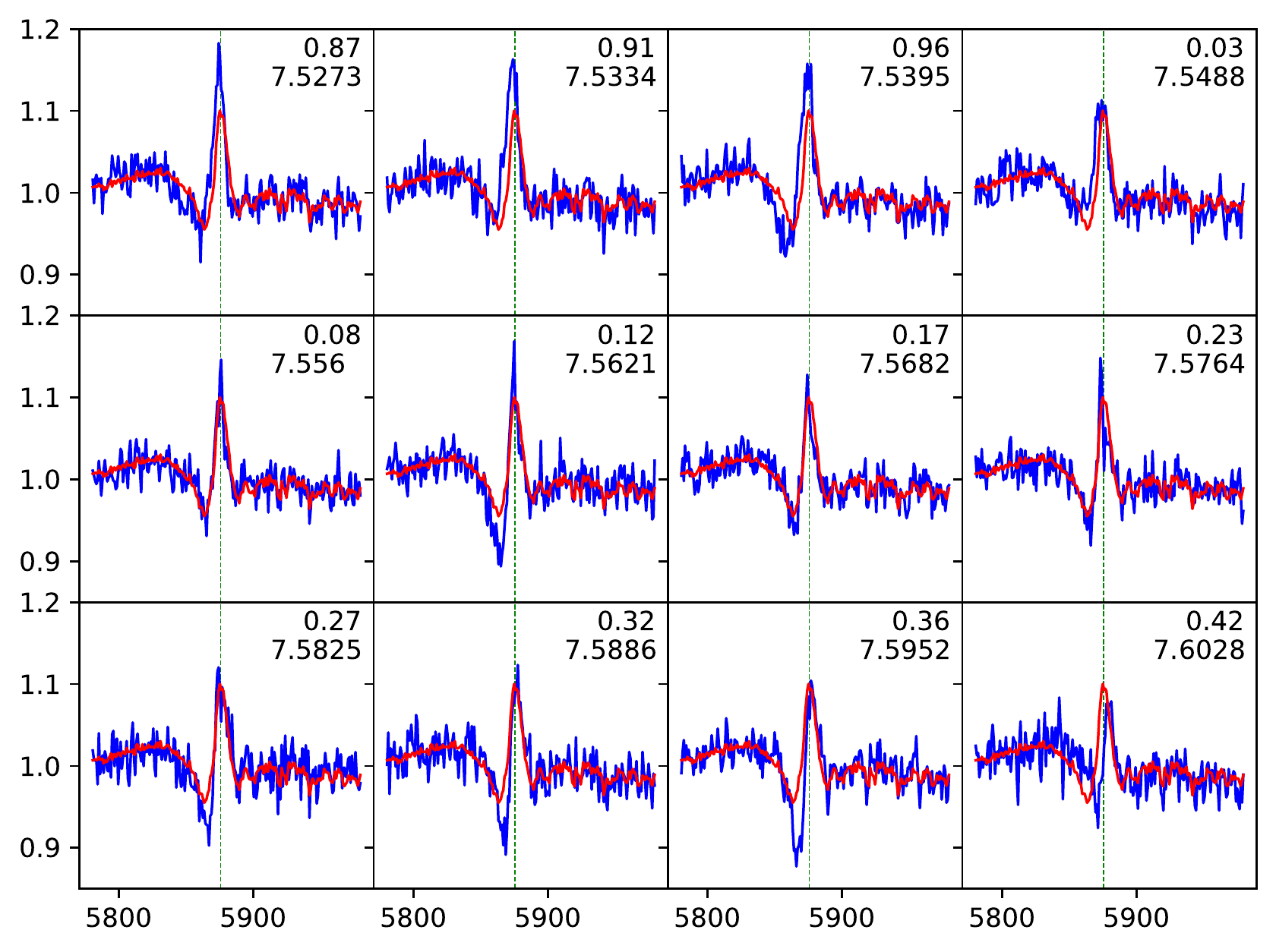}
\caption{\label{fig:spectral evolution}  
Phase resolved spectra centred on H$\alpha$ (top), H$\beta$ (middle) and \ion{He}{i} 5876\,\AA (bottom) from the December INT data. The individual exposures were binned such that each sub-panel in the sequence covers $\simeq8$\,min. The top number in each panel is the orbital phase and the lower number is HJD$-$2\,459\,190. The red line is the time averaged spectrum of the December sample. Note the stochastic variability in the emission lines.}
\end{figure}

\begin{figure*} 
\includegraphics[width=\columnwidth]{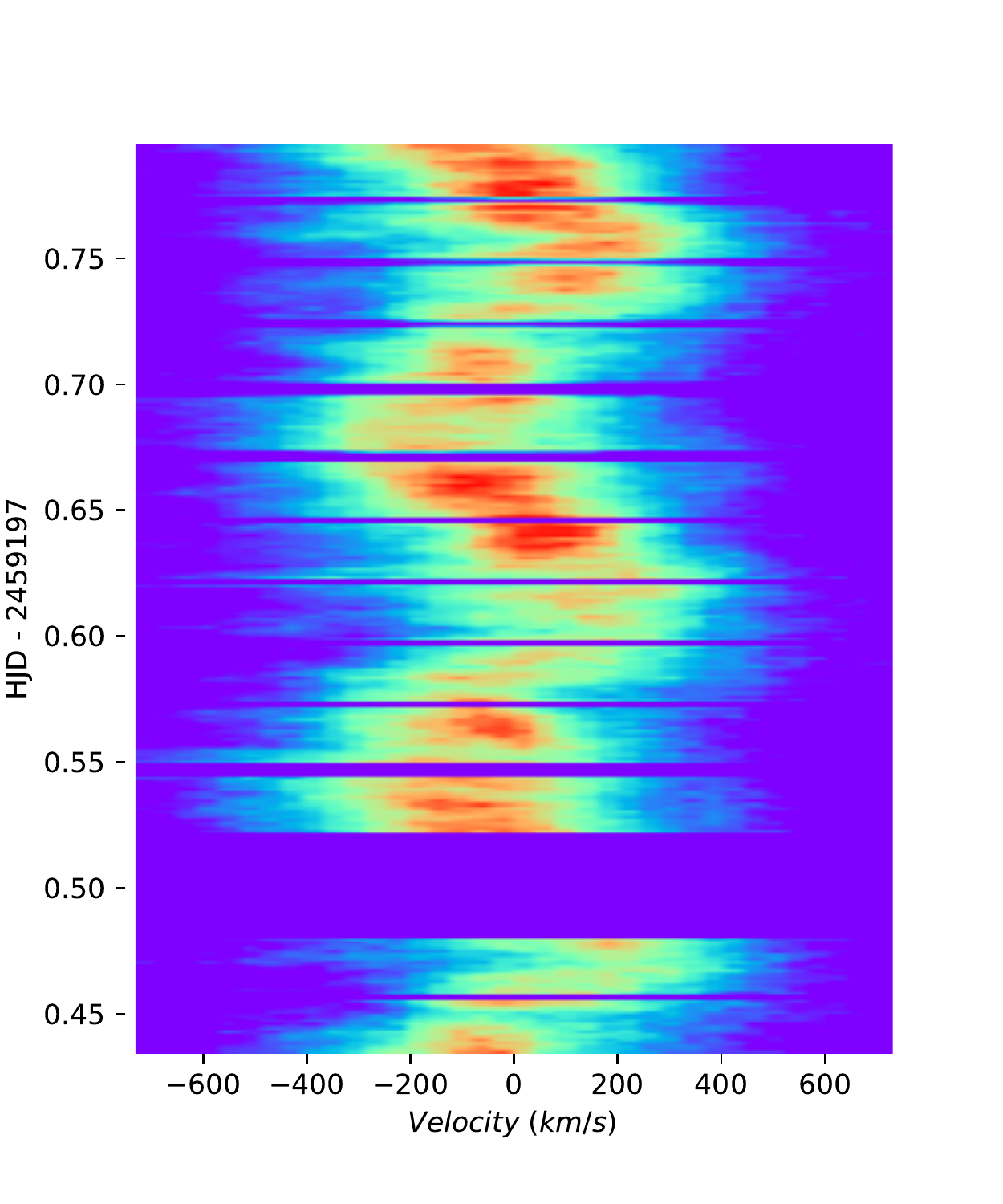}
\includegraphics[width=\columnwidth]{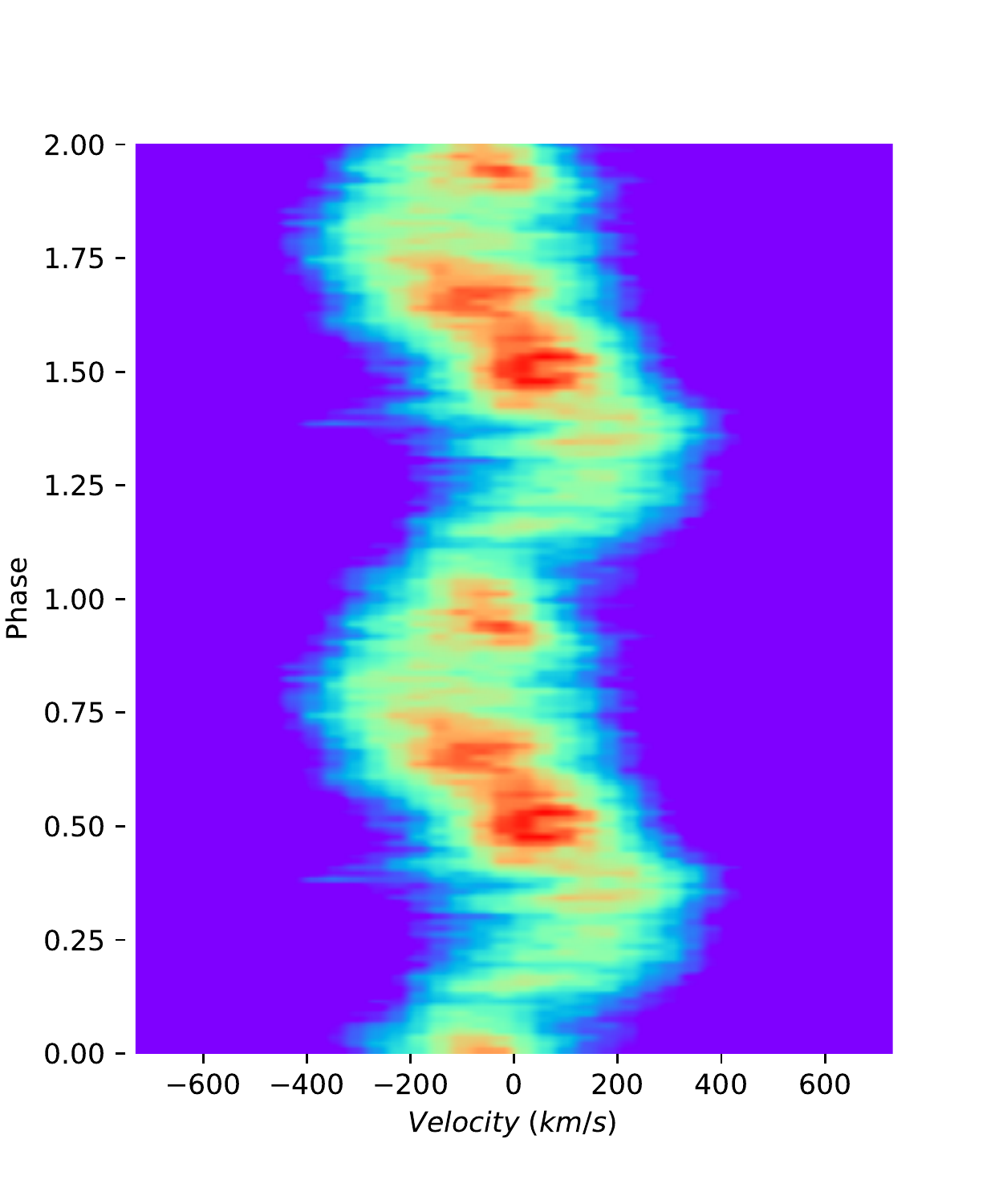}
\caption{\label{fig:trailed_spectrum}  
(left) The trailed spectrum of the H$\alpha$ line in the December sample and (right) the same data phase-folded using the fitted period and epoch. The ``S" shape is clearly visible. Note also the stochastic peaks in the left plot that are less evident in the right hand plot due to averaging and orbital binning. We inspected the trailed spectra of  the earlier observations which provided partial orbital coverage, and did not find repeating orbital behaviour.}
\end{figure*}

\subsection{Spectral variability and detection of a wind outflow}
The high cadence of our INT spectroscopy reveals  rapid variations in the morphology of the emission lines. This is evident in the H$_{\alpha}$ trailed spectrum  (see Fig.\,\ref{fig:trailed_spectrum}) and is particularly striking in the variability of a P\,Cygni-like feature in \Line{He}{I}{5876} (see bottom panel in  Fig.\,\ref{fig:spectral evolution}). These variations appear to be stochastic with no relation to orbital phase or indeed variations in other lines (see Fig.\,\ref{fig:spectral evolution}).

\begin{figure} 
\includegraphics[width=\columnwidth]{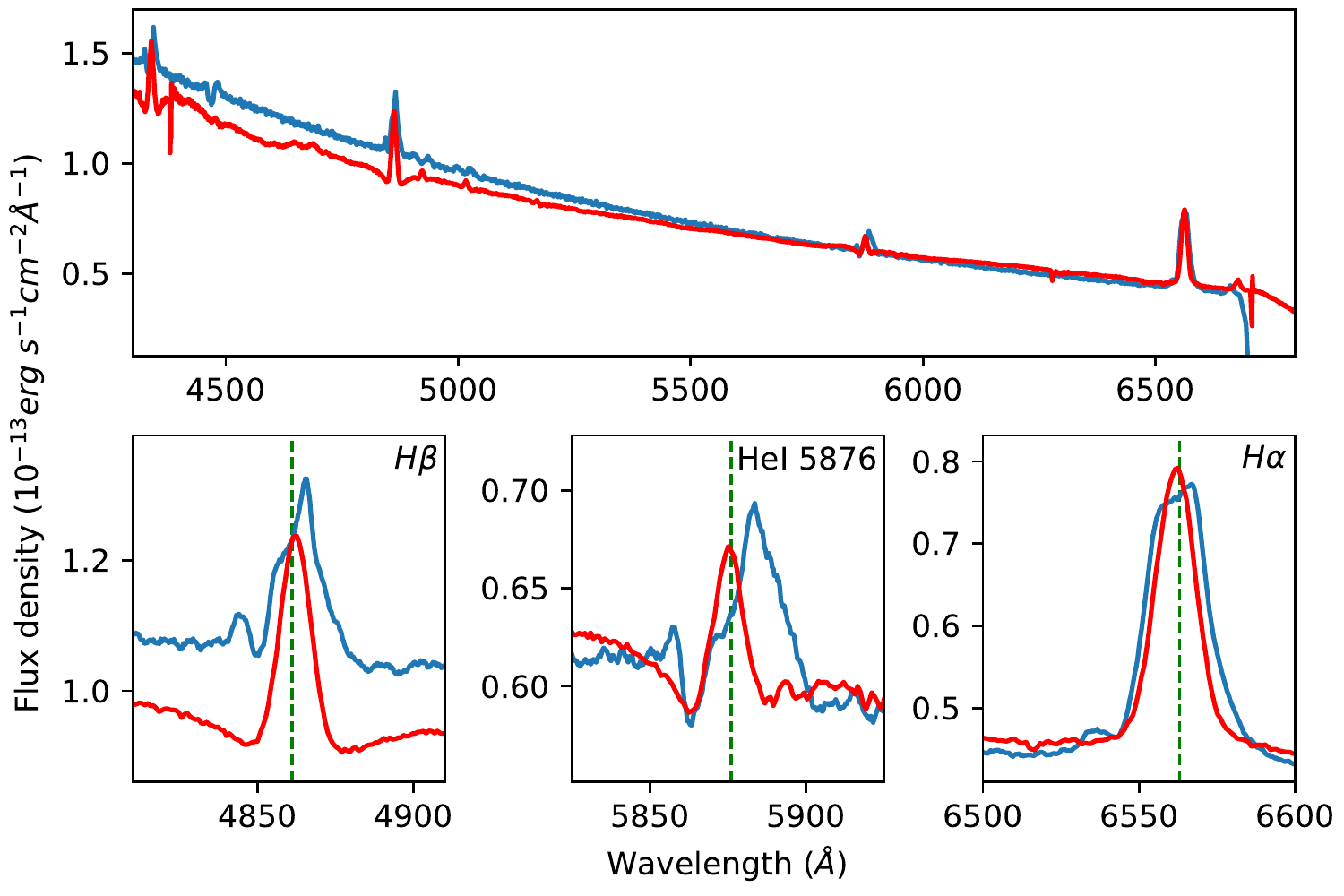}
\caption [.] {\label{fig:pythonmodel} Top panel: the  \textsc{python} model fit (blue) to the observed average spectrum (red). An inclination of 62$\degr$ was found to best reproduce the observed data. Bottom panel: enlargements of the H$\beta$, \Line{He}{I}{5876} and H$\alpha$ lines. }
\end{figure}

\begin{table}
\caption{\label{tab:python} Adopted values used for fitting the \textsc{python} model. See \citet{2015MNRAS.450.3331M} for more details of the model and parameter definitions. }
\centering
\begin{tabular}{ l l c  } 
\hline
\multicolumn {2}{l}{Parameter} & 
\multicolumn {1}{
c}{Value}   \\
\hline
Wind fill factor               &            & 0.008 \\ 

Wind mass transfer rate   & $\dot{M}$\,[$10^{-10}M_{\sun} yr^{-1}$]              & 5.0  \\ 

Mass (white dwarf)               & $M_{1}$\,[$\Msun$]           & 0.83 \\ 
Mass (donor)               & $M_{2}$\,[$\Msun$]           & 0.22 \\
Radius (white dwarf)            & $R_{1}$\, [$\Rsun$]  &0.0122  \\ 
Radius (donor)                  & $R_{2}$\, [$\Rsun$]           & 0.30  \\ 
Inner radius (disc)          & $R_\mathrm{disc,i}$\, [$\Rsun$]           & 0.0244 \\ 
Outer radius (disc)    & $R_\mathrm{disc,o}$\, [$\Rsun$]           & 0.49  \\ 
Temperature (white dwarf)& $T_\mathrm{1}$\, [K]        & 41158\\ 
Temperature (donor)& $T_\mathrm{2}$\, [K]        & 3324\\

Orbital period &$P_{\mathrm{orb}}\,$[hr] 	 &	 $3.2794$   \\

Wind inner radius&$r_{\mathrm{min}}\,$[$R_{1}$] 	&	 $4$   \\ 
Wind outer radius&$r_{\mathrm{max}}\,$[$R_{1}$]	&	 $12$   \\ 

Inner opening angle&$\theta_{\mathrm{min}}\,[^{\circ}]$ & $20.0^{\circ}$ \\
Outer opening angle&$\theta_{\mathrm{max}}\,[^{\circ}]$ & $65.0^{\circ}$\\ 
Angle exponent&$\gamma$ 	&	 $1$    \\ 
Terminal velocity&$v_{\infty}\, [v_{\mathrm{esc}}]$	&	 $3$    \\ 
Wind Acceleration length &R$_\mathrm{v}\,[10^{11}$cm]           & 1.0   \\ 
Acceleration exponent&$\alpha$ 	&	 $1.2$  \\

\hline
\end{tabular}
\end{table}

\section{Discussion} \label{sec:discussion}

\subsection{Winds}
The optical P\,Cygni lines are clear evidence of the presence of winds \citep{2004AJ....128.2420K}. 
Our INT spectroscopy demonstrates  variability within the wind structure on the time scale of minutes~--~probably due to clumping. The lack of a correlation of the wind features with the orbital phase precludes a stable and localised origin within the disc. The blue absorption wing of the \Line{He}{I}{5876} P\,Cygni line is unusually strong compared to other well-studied nova-likes (e.g. \citealt{2004AJ....128.2420K, 2009AJ....137..197K}) and this may be due to a fortuitous inclination. \citet{1998MNRAS.295..595P} finds that slower dense winds are confined to within 45\degr of the orbital plane. The radial velocity component of the winds will therefore be greater in high inclination systems such as ASAS\,J0714+7004. We did not detect the concentration of winds at the superior conjunction of the donor as reported by \citet{2013AJ....145...45H} in BZ\,Cam. 

To explore whether a wind outflow is plausible in the case of  ASAS\,J0714+7004, we have calculated simulated spectra for bi-conical outflow using the Monte Carlo radiative transfer code \textsc{python}  \citep{2002ApJ...579..725L}, which has been used in several previous attempts to model winds in disc-dominated cataclysmic variables at ultraviolet \citep{2010ApJ...719.1932N} and optical wavelengths \citep{2015MNRAS.450.3331M}.    \textsc{python} starts with a parameterised description of a biconical outflow, in this case one originally formulated by \cite{1993ApJ...409..372S}. It then solves for the ionisation structure of the outflow given the nature of the white dwarf and disc, and then simulates the emergent spectrum as a function of the inclination angle\footnote{\textsc{python} is a collaborative open-source project available at \hyperlink{https://github.com/agnwinds/python}{github.com/agnwinds/python}.}.  The \citet{1993ApJ...409..372S} models have a large number of parameters.   For the modelling discussed here, we explored a small fraction of the allowable parameter space, beginning with models that were similar to those described by \citet{2015MNRAS.450.3331M}.  For the models we used parameter values consistent with those derived earlier (Table\,\ref{tab:model}), and varied various of the parameters in Table\,\ref{tab:python}.  The external radiation sources were the white dwarf and the disc; solar abundances were assumed; H and He were treated as multi-level macro-atoms, while the metals were treated in the so-called two-level approximation.  

A comparison of the simulated to the observed spectra for the best model we calculated is shown in Fig.\,\ref{fig:pythonmodel}.  The line strengths and line shapes of the H and He lines are approximately correct.   \Line{He}{I}{5876} shows evidence of a P\,Cygni like feature.  Despite being produced from a wind that arises from the disc at distances between 4 and 12 white dwarf radii, the lines in the simulated spectra and in the observed INT spectra are single peaked.  The reason that the lines are single peaked in the model spectra (at the resolution of the INT spectra) is likely to be due to two effects.  First,   there is emission from fairly far out in the wind which reduces the azimuthal velocity since the flow is designed to conserve angular momentum. In this model, the fact that the P\,Cygni absorption line is $\simeq600\,\mathrm{km\,s^{-1}}$ implies that the interaction with the wind occurs at a considerable distance ($\sim 5-6\,\Rsun$ i.e. $>5$ times the binary separation)  from the system.  Second, the blue edge of a double peaked line tends to be absorbed by material further out in the flow.  With somewhat different parameters one often sees a double H$\alpha$ profile with the blue wing suppressed.  

Although the comparison of model spectra to the observations suggests that this might be a plausible wind model for ASAS\,J0714+7004, and although the model is not that different from those that \cite{2015MNRAS.450.3331M} used to model the spectra of UX\,UMa and RW\,Tri, there are a number of critiques of the model that might be made.  First, while the model spectra resemble the observed spectra, the model spectra are quite sensitive to the inclination angle.  The nature of the simulated spectra change fairly rapidly with inclination angle near $62\degr$ because this is very close to the outer opening angle of the conical flow; at lower angles the P\,Cygni lines grow more prominent; at higher angles the emission lines grow in strength. Changing the outer opening angle to a somewhat lower value, would allow one to produce similar spectra at similarly lower inclination angles. Second, the wind model we use has a mass accretion rate of $10^{-9}\,\Msy$, which compares with the mass accretion rate we have derived at high inclination of  $8.9\times10^{-9}\,\Msy$; this is probably the maximum value that is plausible. Furthermore we have assumed that the wind is clumped, and only occupies three~per cent of the volume. If one preferred a lower mass loss rate, one could have simply assumed that the wind occupied an even smaller fraction of the volume, and obtained a fairly similar spectrum, since the ionisation parameter at any particular point depends on the density of the filled fraction of the wind.  Finally, we note in passing, that there are several \ion{He}{i} lines that appear double peaked in our models, but which are single peaked in the time-averaged spectra. Despite these difficulties, we would suggest that the models indicate that ASAS\,J0714+7004 has a relatively wide-angle outflow with properties that are similar to those in the model, and similar to those in other nova-like systems.

\subsection{Disc structure}
The extensive INT spectroscopy never shows evidence of the double-peaked  emission lines we would expect from a high inclination disc \citep{1986MNRAS.218..761H}. We considered the possibility that ASAS\,J0714+7004 was an SW\,Sex star \citep{2013MNRAS.428.3559D} as this would explain the lack of double-peaked lines even though it did not have the other SW\,Sex defining characteristics (deep continuum eclipses and emission lines that do not share the orbital motion of the white dwarf and exhibit transient absorption features).

However, in contrast to a large fraction of the confirmed SW\,Sex stars, ASAS\,J0714+7004 is not eclipsing and the trailed spectrum shows no evidence of a second peak. A wide flared disc, as proposed to explain the lack of double-peaked lines in SW\,Sex stars \citep{2013MNRAS.428.3559D}, would require a flare angle of over 20\degr\ which is geometrically unlikely. The alternative explanation is that the single-peaked emission lines originate in the wind, as discussed above.

\subsection{Mass transfer rate}

The widely accepted thermal disc instability model \citep{1981A&A...104L..10M, 1996PASP..108...39O} predicts that CVs with a mass transfer rate ($\dot{M}$) below a critical value ($\dot{M}_\mathrm{crit}$) will exhibit outbursts, and such systems are known as dwarf novae. Nova-likes are CVs with $\dot{M}$  above this critical value and are expected to have discs in a steady high-luminosity state. The actual value for $\dot{M}_\mathrm{crit}$ depends on the size of the disc, and hence largely on the orbital period, and is somewhat model dependent  \citep{1986ApJ...305..261S, 2011ApJS..194...28K}. The estimated range of the mass transfer rate for ASAS\,J0714+7004 is typical for that found among nova-likes  (e.g. table 4 from \citealt{2020MNRAS.496.2542H}), and places it above $\dot{M}_\mathrm{crit}$. 

\begin{figure*}
\includegraphics[width=\textwidth]{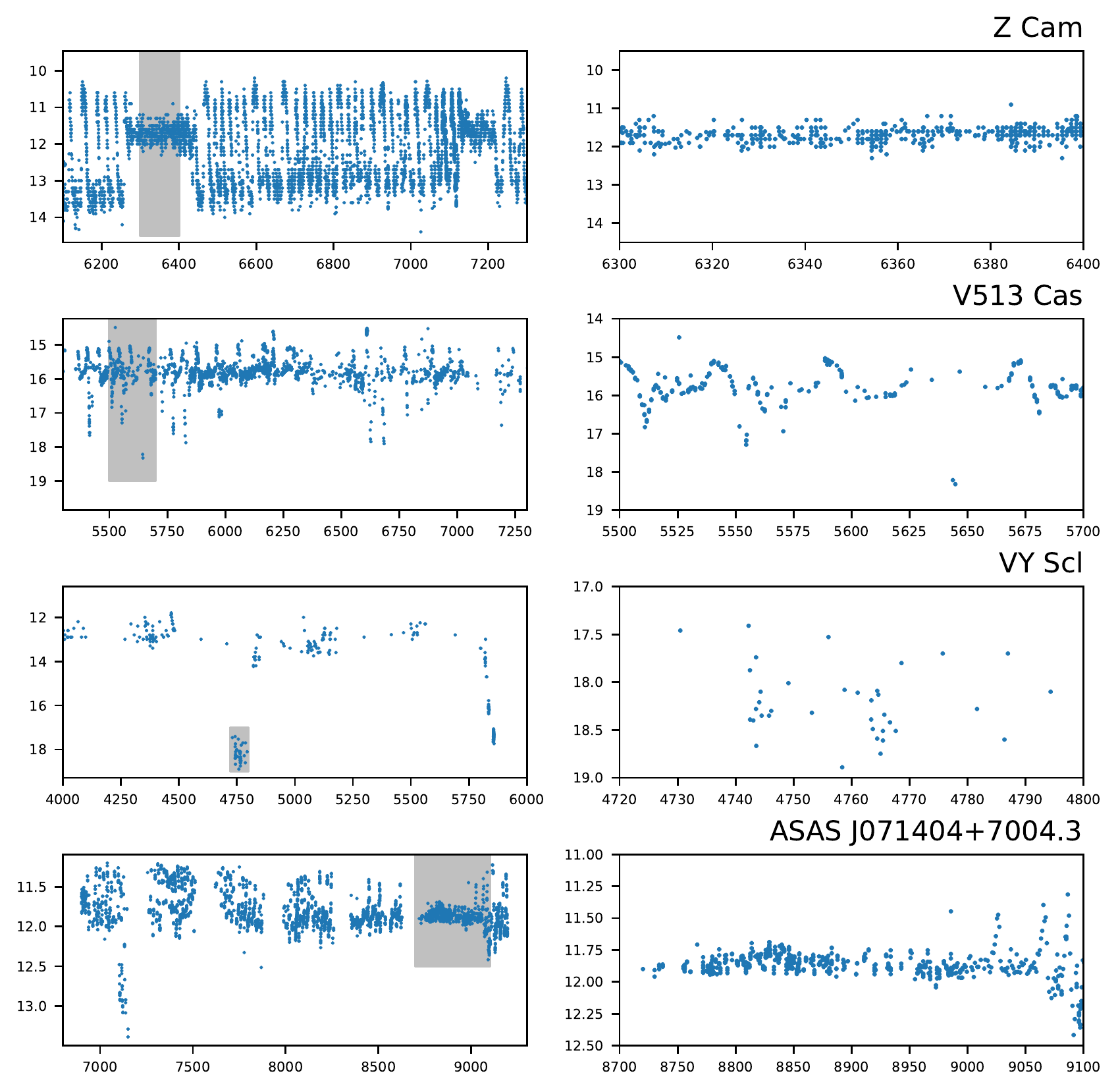}
\caption{Light curves of several typical dwarf nova and nova-like sub-classes. Dates are MJD$-$2\,450\,000. The panels on the right are enlargements of the grey areas in the left panels. The dwarf nova Z\,Cam shows classic ``standstills" of $\simeq100-200$\,d during which there are no outbursts whilst exhibiting large-amplitude outbursts between standstills; standstills are normally terminated by a decline to quiescence. This contrasts with the IW\,And class (of which V513\,Cas is an example)  in which characteristic outbursts occur during standstills. VY\,Scl stars show relatively deep ($2-5$\,mag) low states which can last a few weeks to about a year, and during which small outbursts occur. (Observations from the AAVSO International Database, https://www.aavso.org).} \label{fig:examples}
\end{figure*}

We have used a simple blackbody model for a boundary layer as, in the optical, it has little effect on the fitted value of $\dot{M}$. However this is not the whole story.   \citet{2007AJ....134.1923P} concluded that the standard model needed to be augmented with a modification to the temperature profile of the inner disc, either by increasing the temperature close to the white dwarf or by the addition of a wind or corona component \citep{1994A&A...288..175M}. 
\citet{ 2014ApJ...794...84B} analysed three nova-likes and confirmed this conclusion. These authors suggested that the boundary layer consists of either optically thin regions traversed by Advection-Dominated Accretion
Flows or X-ray corona regions close to the white dwarf but above the disc plane. This then leads to speculation as to why optically thin regions of this sort form in nova-likes but not in dwarf novae. \citet{2019A&A...628A.121N} suggests that the cause is a magnetically controlled zone that builds up due to the persistently ionised disc; dwarf nova outbursts do not last long enough for the field to build up.  

In this context we note that the X-ray spectrum of ASAS\,J0714+7004 is less than one percent of the optical flux which is typical of nova-likes (Figure\,3 in  \citealt{2017PASP..129f2001M}) and clearly shows that the expected 50~per cent of gravitational energy released at the white dwarf boundary  does not escape in the form of X-rays.  ASAS\,J0714+7004 is therefore another example of where the standard model fails to account for this energy.

\subsection{Quasi-periodic outbursts}\label{sec:quasiperiodic}
The photometry in Fig. \ref{fig:photometry} displays quasi-periodic variability of $\simeq1$\,mag on time scales  of $\simeq10-20$\,d, reminiscent of the ``stunted'' outbursts seen in some other nova-likes \citep{1998AJ....115.2527H}. We investigated the possibility of non-orbital periodic variability and found no evidence for this despite a number of apparently regular structures. The \textit{TESS} power spectrum (Fig.\,\ref{fig:tess_20s_power}) clearly confirms the absence of any other periodicities. This quasi-periodic variability is  distinct from lower level and rapid ``flickering" \citep{1992A&A...266..237B} which is evident in the \textit{TESS} light curves (Fig.\,\ref{fig:photometry}, particularly panels (e) and (f)).  The cause of the stunted outbursts remains unclear. One possibility is stochastic changes in the mass transfer rate whilst another is a form of disc instability similar to the cause of dwarf nova outbursts that involves only a small fraction of the disc.  \citet{2001PASP..113..473H} strongly favoured the model of dwarf nova outbursts arising from a disc instability~--~the apparent stunting of the outbursts being caused by dilution from an additional source of illumination within the system.  \citet{2001PASP..113..473H} also argued that the presence of stunted outbursts is not correlated with $\dot{M}$ which suggests that they are not due to those nova-likes displaying them being close to $\dot{M}_\mathrm{crit}$. Inspecting the long-term light curve of ASAS\,J0714+7004 (top panel in Fig.\,\ref{fig:photometry}) shows that the (dis)appearance of the quasi-periodic variability does indeed not correlate strongly with the average brightness of the system. 

\citet{2018AJ....155...61R} propose a contrary view based on observational evidence of UU\,Aqr that about half of the stunted outbursts coincided with increased hot spot emissions and hence higher $\dot{M}$. Subsequently  \citet{2019ApJ...876..152S} studied an eclipsing nova-like (AC\,Cnc) and found that the stunted outbursts were eclipsed by the disc suggesting that either the outbursts originated on the facing hemisphere of the donor or were due to changes in the disc. 

Closer investigation of an outburst observed at two minute cadence by \textit{TESS}  (see Fig.\,\ref{fig:photometry}, panel (g)) shows a gradual rise and fall typical of an ``inside out'' dwarf nova. The \textit{Swift} ultraviolet observations show a similar outburst with AAVSO optical observations also peaking at the same time. Given that the bulk of the disc must be ionised to sustain the high $\dot{M}$ we speculate that there is an outer part of the disc that occasionally becomes sufficiently cool for H to recombine, and that this outburst starts at the inner edge of this area and subsequently spreads outwards. As such, we conclude that the quasi-periodic variability seen in ASAS\,J0714+7004 is likely related to the standard thermal disc instability acting in a small outer part of the disc, leaving the inner disc in a steady hot state which providies the ``background'' luminosity that \citet{2001PASP..113..473H} required to dilute the amplitude of the outbursts.

\subsection{Nova-likes and dwarf novae~--~secular changes in mass transfer rate} 
The observed temporary change to a low state in 2015 ($\mathrm{HJD} - 2450000 =  7150$) is interesting as it leads to a potential explanation for the under-representation of dwarf novae with orbital periods between $\simeq3-4$\, hours~--~just above the so-called ``period gap" \citep{1986ApJ...305..261S}. 

\citet{2011ApJS..194...28K} showed that populations of dwarf novae and nova-likes coexist at these orbital periods and so the direct association between orbital period and mass transfer rate breaks down in this region. Our observation supports the idea that CVs can change back and forth between dwarf novae and nova-likes more frequently (a few years)~--~particularly when mass transfer rates are close to $\dot{M}_\mathrm{crit}$. Nova-likes are brighter (by $\simeq5$\,mag) than dwarf novae and hence there will be a large selection effect leading to the under-representation of dwarf novae at such periods.

In the light of our observations it is interesting to compare ASAS\,J0714+7004 with known sub-classes of nova-likes (see Fig.\,\ref{fig:examples}). 

Z\,Cam stars (e.g. \citealt{2002A&A...384L...6S}) are dwarf novae that experience occasional standstills of constant magnitude of $1-1.5$\,mag below maximum usually following an outburst. The remainder of the time Z\,Cam stars exhibit frequent dwarf novae commencing $\simeq1$\,mag below the standstill level and peaking above it. The bottom right light curve in Fig.\,\ref{fig:examples} shows ASAS\,J0714+7004 has a high-level standstill but the amplitude of the outbursts at other times is far lower than Z\,Cam. 

The IW\,And stars \citep{2014A&A...569A..48H} form a sub-category of Z\,Cam (see Fig.\,\ref{fig:examples}) and show continual outbursts (not unlike dwarf novae) during the high state (i.e. during the ``standstill"). Again this is not consistent with our light curve except for our low state in 2015.

The VY\,Scl stars \citep{2018A&A...617A..16S} are nova-likes that show occasional $\sim3-6$\,mag low states similar to ASAS\,J0714+7004 in 2015 and typically have orbital periods of $3-4$\,h. Whilst the cause of the low state is not well understood it is clear that the previously hot steady-state disc will continue to deposit its mass onto the white dwarf until it is exhausted~--~potentially causing dwarf nova outbursts which we do see (panel (b) in Fig.\,\ref{fig:photometry}). Based on these observations we conclude that ASAS\,J0714+7004 is a VY\,Scl star.


\section {Conclusions} \label{sec:conclusions}
Our study of ASAS\,J0714+7004 has shown it to be a bright  and nearby nova-like CV. It exhibits the low states characteristic of the VY\,Scl sub-class. We have obtained a reliable estimate of the orbital period and estimated the mass transfer rate and inclination  (see Fig.\,\ref{tab:ephemera}). We have identified strong wind signatures that are highly variable~--~coming and going in a few minutes in veritable gusts, and uncorrelated with orbital phase. Because of its brightness and proximity, this system is an excellent target for future detailed studies of its accretion disc and wind outflow.

\section {Acknowledgements}
We thank Professor David Buckley for completing a detailed review and helpful comments.

We gratefully acknowledge the photometric observations of ASAS\,J0714+7004 obtained in 2020 and 2021 by Tam\'as Tordai (Hungary), Stephen Brincat (Malta), Kenneth Menzies (USA), Charles Galdies (Malta), Sjoerd Dufoer (Belgium), Thanasis Papadimitriou (Greece), Richard Sabo (USA), David Boyd (UK), Molly Wakeling (USA) and Steve Boemer (USA) which we retrieved from the AAVSO International Database.
Based on observations made with the Isaac Newton Telescope operated on the island of La Palma by the Isaac Newton Group of Telescopes in the Spanish Observatorio del Roque de los Muchachos of the Instituto de Astrof\'isica de Canarias. The IDS spectroscopy was obtained as part of P9/C48/N1 and N1/C48/P9. 
This paper includes data collected with the TESS mission, obtained from the MAST data archive at the Space Telescope Science Institute (STScI). Funding for the TESS mission is provided by the NASA Explorer Program. STScI is operated by the Association of Universities for Research in Astronomy, Inc., under NASA contract NAS 5–26555.
We greatly appreciate the use of the spectres python code to rebin spectra  \citep{2017arXiv170505165C}.
This work made use of data supplied by the UK Swift Science Data Centre at the University of Leicester.
We acknowledge with thanks the processing of the UVOT data by Paul Kuin (Mullard Space Science Laboratory, University College London).
We acknowledge that this work made use of the Tych0-2 catalog, 
This work has made use of data from the European Space Agency (ESA) mission \textit{ Gaia} (\url{https://www.cosmos.esa.int/gaia}), processed by the \textit{Gaia} Data Processing and Analysis Consortium (DPAC, \url{https://www.cosmos.esa.int/web/gaia/dpac/consortium}). Funding for the DPAC has been provided by national institutions, in particular the institutions participating in the \textit{Gaia} Multilateral Agreement.
This publication makes use of data products from the Two Micron All Sky Survey, which is a joint project of the University of Massachusetts and the Infrared Processing and Analysis Center/California Institute of Technology, funded by the National Aeronautics and Space Administration and the National Science Foundation.
This publication makes use of data products from the Wide-field Infrared Survey Explorer, which is a joint project of the University of California, Los Angeles, and the Jet Propulsion Laboratory/California Institute of Technology, funded by the National Aeronautics and Space Administration.
The Pan-STARRS1 Surveys (PS1) and the PS1 public science archive have been made possible through contributions by the Institute for Astronomy, the University of Hawaii, the Pan-STARRS Project Office, the Max-Planck Society and its participating institutes, the Max Planck Institute for Astronomy, Heidelberg and the Max Planck Institute for Extraterrestrial Physics, Garching, The Johns Hopkins University, Durham University, the University of Edinburgh, the Queen's University Belfast, the Harvard-Smithsonian Center for Astrophysics, the Las Cumbres Observatory Global Telescope Network Incorporated, the National Central University of Taiwan, the Space Telescope Science Institute, the National Aeronautics and Space Administration under Grant No. NNX08AR22G issued through the Planetary Science Division of the NASA Science Mission Directorate, the National Science Foundation Grant No. AST-1238877, the University of Maryland, Eotvos Lorand University (ELTE), the Los Alamos National Laboratory, and the Gordon and Betty Moore Foundation.
We gratefully acknowledge NASA’s support for construction, operation, and science analysis for the GALEX mission, developed in cooperation with the Centre National d’Etudes Spatiales of France and the Korean Ministry of Science and Technology.
Support for the ATLAS survey was provided by NASA grant NN12AR55G under the guidance of Lindley Johnson and Kelly Fast. 
We have made use of the ROSAT Data Archive of the Max-Planck-Institut für extraterrestrische Physik (MPE) at Garching, Germany.
This research has made use of the XSPEC tool obtained from the High Energy Astrophysics Science Archive Research Center (HEASARC), a service of the Astrophysics Science Division at NASA/GSFC and of the Smithsonian Astrophysical Observatory's High Energy Astrophysics Division."
BTG and TRM were Leverhulme Research Fellowships 
and the UK STFC grant ST/T000406/1.

\section {Data availability}
The INT spectroscopy will be available via the data archives of the ING after the expiry of the default proprietary period. All the time-series photometry is available via the archives of the AAVSO and \textit{TESS} mission. 



\bibliographystyle{mnras}
\bibliography{refs} 

\clearpage


\clearpage
\appendix

\section{From \textit{Gaia} DR2 to EDR3: improved precision and overall changes due to variability}\label{sec:appendix}

\begin{figure*}
\includegraphics[width=0.7\textwidth]{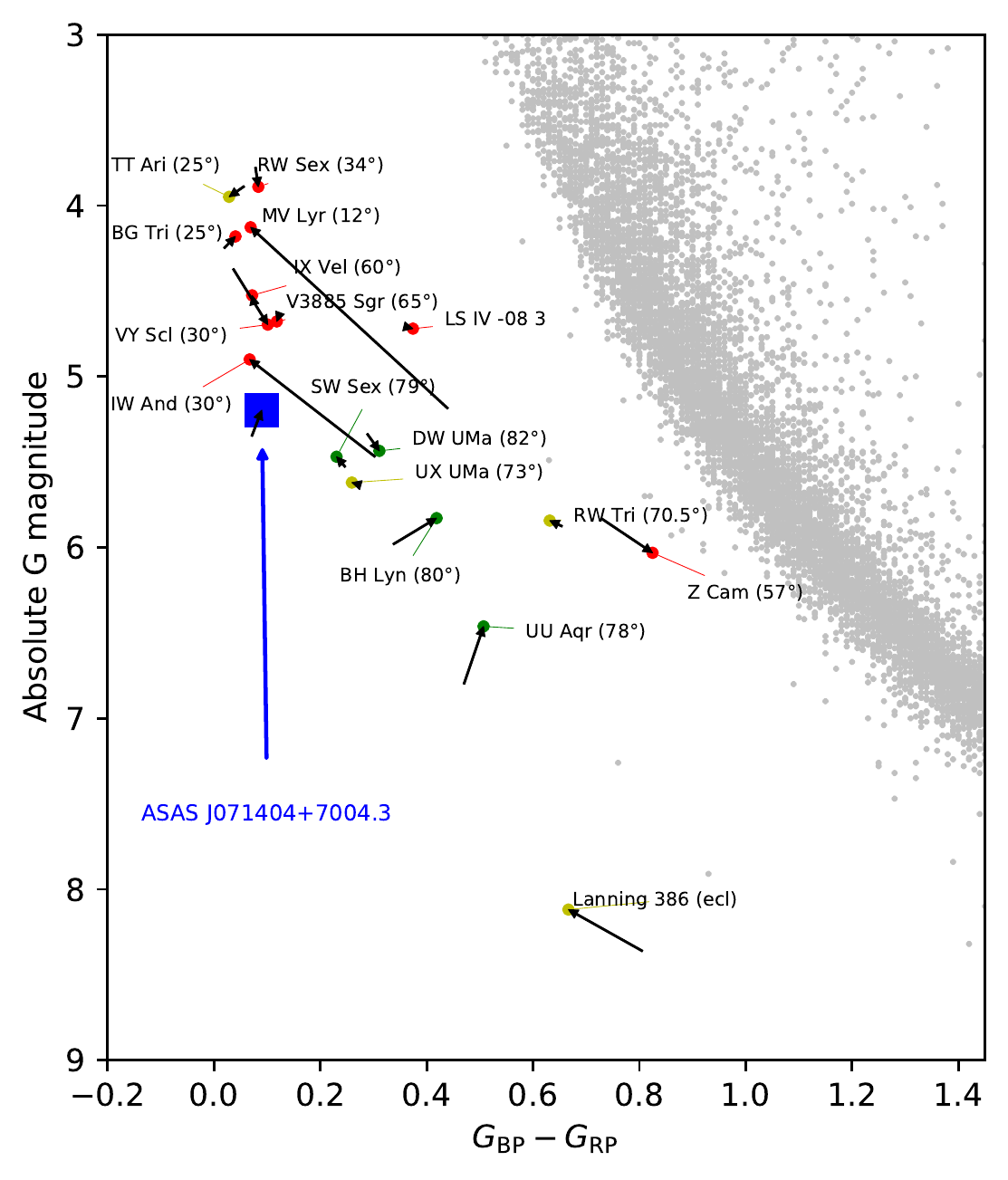}
\caption{ Selected nova-likes in the \textit{Gaia} EDR3 HRD showing the changes (black arrows) in absolute magnitude and colour between \textit{Gaia} DR2 and EDR3. Whilst some minor changes are due to refinement in \textit{Gaia} astrometry and photometry the major changes are due to transitory changes in the observed light curve. MV\,Lyr is a good example of this where it was mostly in a low state during the 162 observations contained in DR2 but in a high state during the remaining 107 observations in EDR3) see Fig.\,\ref{fig:MV Lyr})   }  \label{fig:edr3 changes}
\end{figure*}

\begin{figure*}
\includegraphics[width=\textwidth]{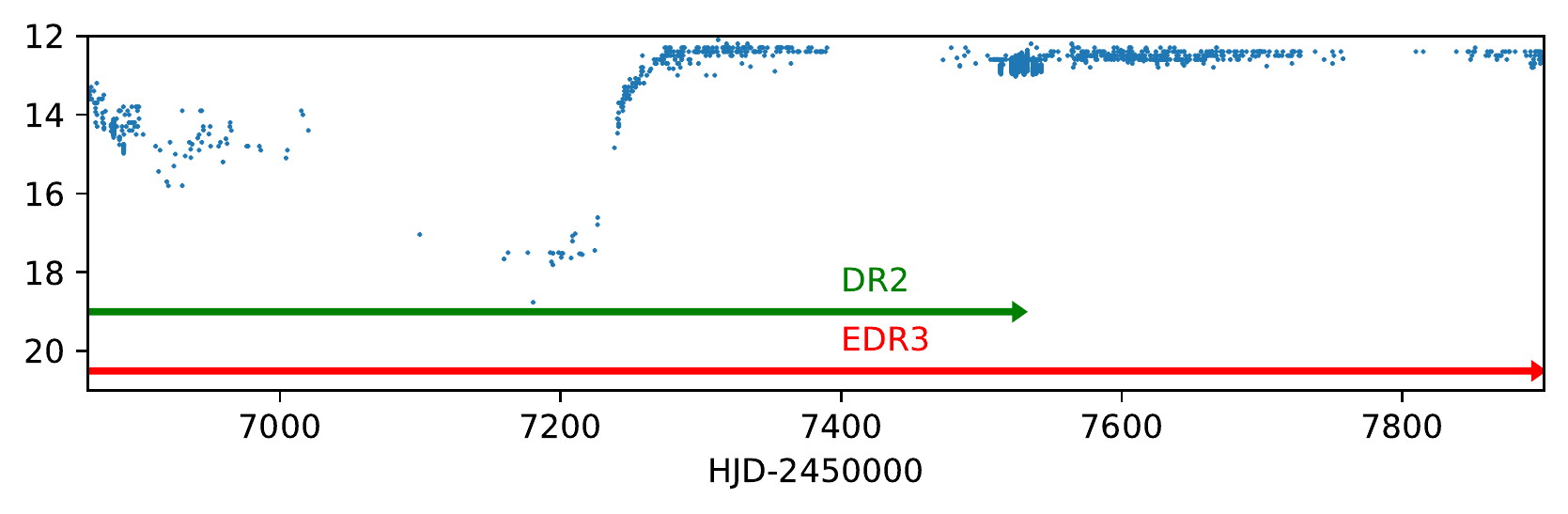}
\caption
{
Long-term light curve of MV\,Lyr, obtained from the AAVSO data base. The arrows indicate the date range covered by the observations contained in \textit{Gaia} DR2 and EDR3, illustrating why the absolute magnitude of this system dropped significantly in EDR3 compared to DR2.}  \label{fig:MV Lyr}
\end{figure*}

\textit{Gaia} EDR3 is a complete re-release of the astrophysical and photometric data and inevitably the absolute magnitudes and colours of any source are expected to differ somewhat with respect to DR2. For the set of nova-likes shown in Fig.\,\ref{fig:HR_diagram}, the uncertainties of the $G$, $G_\mathrm{BP}$, $G_\mathrm{RP}$ magnitudes and the parallaxes were   0.0101, 0.0355, 0.0308 and 0.0102 respectively in EDR3 compared to 0.0144, 0.0423, 0.0346 and 0.0216 in DR2.   Fig\,\ref{fig:edr3 changes} shows the corresponding changes in the \textit{Gaia} HRD as black arrows. In most cases, these changes are quite small but one system (MV\,Lyr) stands out as its absolute magnitude decreased by $\simeq 1$. The reason is an important reminder that nova-likes are variable stars and \textit{Gaia} EDR3 essentially computes a weighted average of measurements taken between 25 July 2014 and 28 May 2017 (whereas DR2 data collection stopped on 23 May 2016).  Fig\,\ref{fig:MV Lyr} shows that MV\,Lyr was in a prolonged low state for about 2/3 of the DR2 observations, whereas EDR3 sampled the system for a substantially longer period in the high state~--~explaining the large change in $G$, $G_\mathrm{BP}$, and $G_\mathrm{RP}$.


\bsp	
\label{lastpage}
\end{document}